\documentclass[graybox]{svmult}

\usepackage{mathptmx}       
\usepackage{helvet}         
\usepackage{courier}        
\usepackage{type1cm}        
\usepackage{makeidx}         
\usepackage{graphicx}        
\usepackage{multicol}        
\usepackage[bottom]{footmisc}
\usepackage{amssymb,amsfonts,amsmath,mathrsfs}
\usepackage{lscape}
\usepackage[figuresright]{rotating}

\def\X{{\mathscr{X}}}
\def\G{{\mathscr{G}}}

\DeclareMathOperator*{\argmax}{argmax}
\DeclareMathOperator*{\argmin}{argmin}
\newcommand*\diff{\mathop{}\!\mathrm{d}}

\newtheorem{observation}{Observation}

\makeindex             

\makeindex
\begin{document}

\title*{Advances in Computational Methods for Phylogenetic Networks in the Presence of Hybridization}
\titlerunning{Computational Methods for Phylogenetic Networks}
\author{R. A. Leo Elworth, Huw A. Ogilvie, Jiafan Zhu,  and Luay Nakhleh}
\institute{R. A. Leo Elworth \at Rice University \email{r.a.leo.elworth@rice.edu}
\and Huw A. Ogilvie \at Rice University \email{huw.a.ogilvie@rice.edu}
\and Jiafan Zhu \at Rice University \email{jiafan.zhu@rice.edu}
\and Luay Nakhleh \at Rice University \email{nakhleh@rice.edu}}
%
%
\maketitle

\abstract{Phylogenetic networks extend phylogenetic trees to allow for modeling reticulate evolutionary processes such as hybridization. 
 They take the shape of a rooted, directed, acyclic graph, and when parameterized with evolutionary parameters, such as divergence 
 times and population sizes, they form a generative process of molecular sequence evolution. Early work on computational methods for phylogenetic network inference focused exclusively on reticulations and sought 
 networks with the fewest number of reticulations to fit the data. 
 As processes such as incomplete lineage sorting (ILS) could be at play concurrently with 
 hybridization, work in the last decade has shifted to computational approaches for phylogenetic network inference in the presence of ILS. In such 
 a short period, significant advances have been made on developing and implementing such computational approaches. In particular, parsimony, likelihood, and Bayesian methods have been devised for estimating phylogenetic networks and associated parameters using estimated gene trees as data. Use of those inference methods has been augmented with statistical tests for specific hypotheses of hybridization, like the $D$-statistic. Most recently, Bayesian approaches for inferring 
 phylogenetic networks directly from sequence data were developed and implemented. In this chapter, we survey such advances and discuss model assumptions as 
 well as methods' strengths and limitations. We also discuss 
  parallel efforts in the population genetics community aimed at inferring similar structures. Finally, we highlight major directions for future research in this area. 
}

 \section{Introduction}
 Hybridization is often defined as reproduction between members of genetically distinct populations \cite{barton1985analysis}. This process could occur 
 in various spatial contexts, and could have impacts on speciation and differentiation \cite{Arnold97,Barton01,Mallet05,Mallet07,Riesberg97,abbott2013hybridization,mallet2016reticulated}. 
Furthermore, increasing evidence as to the adaptive role of hybridization has been documented, for example, in humans \cite{racimo2015evidence}, 
 macaques \cite{stevison2009divergence,bonhomme2009assessing,osada2010ancient}, mice \cite{Song20111296,LiuEtAl15}, butterflies \cite{Heliconius2012,zhang2016genome}, and mosquitoes \cite{fontaine2015,WenEtAl16}.
 
\begin{figure}[!ht]
  \begin{center}
  \includegraphics[width=4.5in]{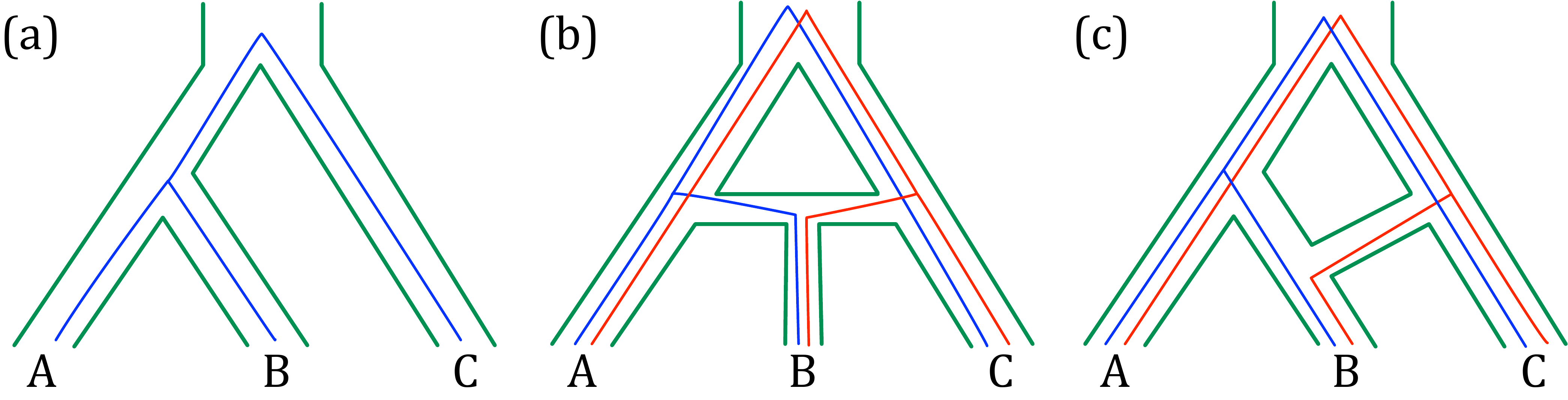}
  \caption{{\bf Hybrid speciation and introgression.} (a) A phylogenetic tree on three taxa, A, B, and C, and a gene tree within its 
  branches. Genetic material is inherited from ancestors to descendants and it is expected that loci across the genome would have the shown gene tree. (b) A hybrid speciation scenario depicted by a phylogenetic network, 
  where B is a hybrid population that is distinct from its parental species. Shown within the branches of the network are two gene trees, both of which are assumed to be very common 
  across the genome. (c) An introgression scenario. Through hybridization and backcrossing, genetic material from (an ancestor of) C is incorporated into the genomes of individuals in (an ancestor of) B. The introgressed 
  genetic material would have the gene tree shown in red, and the majority of loci in B's genomes would have the gene tree shown in blue. Incomplete lineage sorting would complicate all three 
  scenarios by giving rise to loci with other possible gene trees and by changing the distribution of the various gene trees.\label{fig:gtstsn}}
  \end{center}
  \end{figure}

 Hybridization is ``generically" 
 used to contain two different processes: hybrid speciation and introgression \cite{folk2018new}. In the case of hybrid speciation, a new population made of the 
 hybrid individuals forms as a separate and distinct lineage from either of its two parental populations.\footnote{In this chapter, we do not make a distinction between {\em species}, {\em population}, 
 or {\em sub-population}. The modeling assumptions and algorithmic techniques underlying all the methods we describe here neither require nor make use of such a distinction.}  Introgression, or introgressive hybridization, on the other 
 hand, describes the incorporation of genetic material into the genome of a population via interbreeding and backcrossing, yet without creating a new population
 \cite{harrison2014hybridization}. As Harrison and Larson noted \cite{harrison2014hybridization}, introgression is a relative term: alleles at some loci introgress with
 respect to alleles at other loci within the same genomes. From a genomic perspective, and as the basis for detection of hybridization, the general view is that in the case 
 of hybrid speciation, regions derived from either of the parental ancestries of a hybrid species would be common across the genomes, whereas in the case of introgression, regions derived from introgression 
 would be rare across the genomes \cite{folk2018new}. Fig. \ref{fig:gtstsn} illustrates both hybridization scenarios. 

 A major caveat to the aforementioned general view is that, along with hybridization, other evolutionary processes could also be at play, which significantly 
 complicates the identification of hybrid species and their parental ancestries. Chief among those processes are incomplete lineage sorting (ILS) and 
 gene duplication and loss. Indeed, various studies have highlighted the importance of accounting for ILS when attempting to detect hybridization based 
 on patterns of gene tree incongruence \cite{Heliconius2012,Marcussen14,YuEtAl14,fontaine2015,clark2015conundrum,racimo2015evidence,WenEtAl16,zhang2016genome,pease2016phylogenomics}.
  Furthermore, gene duplication and loss are very common across all branches of the Tree of Life. While the main focus of this chapter is on modeling and inferring 
  hybridization, a discussion of how ILS is accounted for is also provided since recent developments have made great strides in modeling hybridization and ILS 
  simultaneously. While signatures of gene duplication and loss are ubiquitous in genomic data sets, we do not include a discussion of these two processes in this 
  chapter since methods that account for them in the context of phylogenetic networks are currently lacking. 
  
 When hybridization occurs, the evolutionary history of the set of species is best modeled by a {\em phylogenetic network}, which extends the phylogenetic 
 tree model by allowing for ``horizontal" edges to denote hybridization and to facilitate modeling bi-parental inheritance of genetic material. Fig. \ref{fig:gtstsn}
 shows two phylogenetic networks that model hybrid speciation and introgression. It is very important to note, though, that from the perspective of existing models, 
 both phylogenetic networks are topologically identical. This issue highlights two important issues that must be thought about carefully when interpreting a phylogenetic network. 
 First, neither the phylogenetic network nor the method underlying its inference distinguish between hybrid speciation and introgression. This distinction is a matter 
 of interpretation by the user. For example, the phylogenetic network in Fig. \ref{fig:gtstsn}(c) could be redrawn, without changing the model or any of its 
 properties, so that the introgression is from (an ancestor of) A to (an ancestor of) B, in which case the ``red" gene tree would be expected to appear with much higher 
 frequency than the ``blue" gene tree. In other words, the way a phylogenetic network is drawn could convey different messages about the evolutionary history that is 
 not inherent in the model or the inference methods. This issue was importantly highlighted with respect to data analysis in \cite{WenEtAl16} (Figure 7 therein). 
  Second, the phylogenetic network does not by itself encode any specific backbone species tree that introgressions could be interpreted with respect to. This, too, is 
 a matter of interpretation by the user. This is why, for example, Clark and Messer \cite{clark2015conundrum} recently argued that ``perhaps we should dispense with the tree and acknowledge that these 
  genomes are best described by a network." Furthermore, recent studies demonstrated the limitations of inferring a species tree ``despite hybridization" 
  \cite{solis2016inconsistency,ZhuYuNakhleh16}. 

\begin{figure}[!ht]
  \begin{center}
  \includegraphics[width=5.5in]{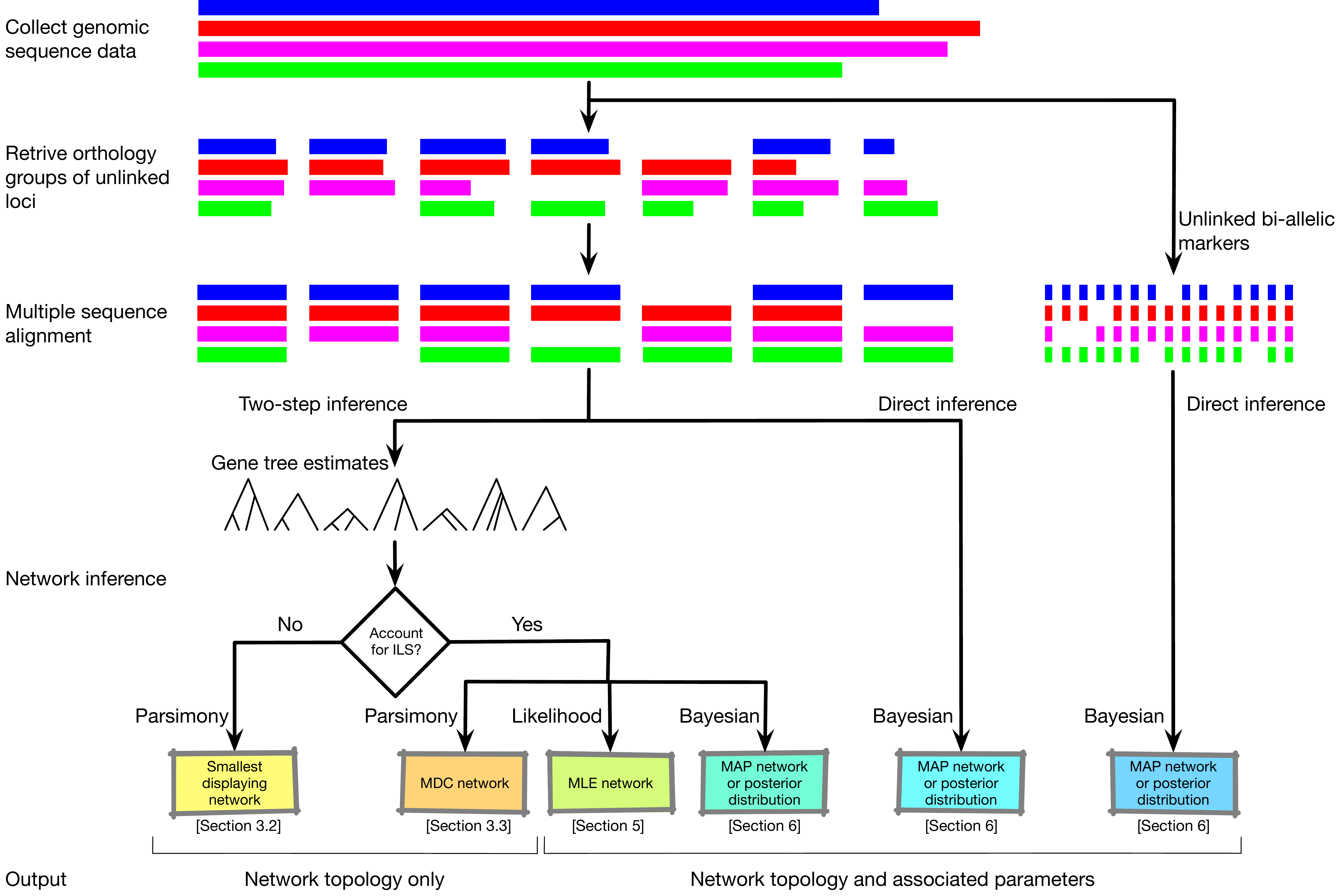}
  \caption{{\bf Phylogenetic network inference process and approaches.} The process of phylogenetic network inference starts with collecting the genomic 
  data and identifying the orthology groups of unlinked loci. Multiple sequence alignments or single bi-allelic markers corresponding to the unlinked loci are 
  then obtained; phylogenetic network inference methods use one of these two types of data. In two-step inference methods, gene trees are first estimated for the 
  individual loci from the sequence alignment data, and these gene tree estimates are used as the input data for network inference. If incomplete lineage sorting (ILS) 
  is not accounted for, a smallest displaying network of the gene tree estimates is sought. If ILS is accounted for, parsimony inference based on the MDC (minimizing 
  deep coalescences) criterion, a maximum likelihood estimate (MLE), a maximum a posteriori (MAP) estimate, or samples of the posterior distribution can be 
  obtained. In the direct inference approach, whether based on sequence alignment or bi-allelic marker data, a MAP estimate or samples of the posterior distribution 
  can be obtained directly from the data. The two parsimony methods consider only the topologies of the gene tree estimates as input (i.e., they ignore gene tree branch lengths) and return as output phylogenetic network topologies. The likelihood and Bayesian methods that take gene tree estimates as input can operate on gene tree topologies alone or gene trees with branch lengths as well. Both methods estimate phylogenetic network topologies along with branch lengths (in coalescent units) and inheritance probabilities. The direct inference methods estimate the phylogenetic network along with its associated parameters. \label{fig:methods}}
  \end{center}
  \end{figure}

  With the availability of data from multiple genomic regions, and increasingly often from whole genomes, a wide array of methods for inferring species trees, mainly based on
  the multispecies coalescent (MSC) model \cite{Degnan09}, have been developed \cite{LiuEtAl09,Nakhleh2013,NYAS:NYAS12747}. Building on these methods,
  and often extending them in novel ways, the development of computational methods for inferring phylogenetic networks from genome-wide data has made great strides in recent years. 
  Fig. \ref{fig:methods} summarizes the general approaches that most phylogenetic network inference methods have followed in terms of the data they utilize, the 
  model they employ, and the inferences they make. 

The overarching goal of this chapter is to review the existing methods for inferring phylogenetic networks in the presence of hybridization,\footnote{We emphasize hybridization (in eukaryotic species) here since processes such as horizontal gene transfer in microbial organisms result in reticulate evolutionary histories, but the applicability of methods we describe in this chapter has not been investigated or explored in such a domain.} describe their strengths and limitations, and highlight major directions for future research 
in this area. All the methods discussed hereafter make use of multi-locus data, where a  locus in this context refers to a segment of genome present across the individuals and species sampled for a given study and related through common descent. A locus can be of varying length, coding or non-coding, and can be either functional or non-functional. Therefore, the use of the term ``gene trees" is only 
historical; we use it to mean the evolutionary history of an individual locus, regardless of whether the locus overlaps with a coding region or not. Care must be taken with increasingly long loci spanning hundreds or thousands of contiguous basepairs, however, as many methods assume a locus has not been affected by recombination. 

Multi-locus methods are fairly popular because the model fits several types of reduced representation genomic data sets commonly generated to study biological systems. Reduced representation refers to capturing many segments scattered throughout a genome, but only covering a fraction of the total genome sequence \cite{Good2011}.
 Reduced representation data sets which have been used with multi-locus methods include RAD-seq and genotyping by sequencing (GBS), which capture loci of roughly 100 bp associated with palindromic restriction enzyme recognition sites \cite{doi:10.1093/sysbio/syx062}. Another family of techniques often applied to studies of deeper time scales, sequence capture, extracts conserved sequences using probes complementary to targetted exons or ultraconserved elements \cite{GRUMMER2018}. Sequence capture can also be performed \textit{in silico} when whole genomes are available \cite{JarvisEtAl14}.
 
 The rest of the chapter is organized as follows. We begin in Section 2 by defining terminology for the non-biologist, and give a very brief review of phylogenetic trees and their likelihood. 
 In Section 3, we describe the earliest, and simplest from a modeling perspective, approaches to inferring parsimonious phylogenetic networks 
 from gene tree topologies by utilizing their incongruence as the signal for hybridization. To account for ILS, we describe in Section 4 the multispecies network coalescent, or MSNC, which is the core model 
 for developing statistical approaches to phylogenetic network inference while accounting for ILS simultaneously with hybridization. In Sections 5 and 6 we describe the maximum likelihood and Bayesian methods 
 for inferring phylogenetic networks from multi-locus data. In Section 7, we briefly discuss an approach aimed at detecting hybridization by using phylogenetic invariants. This approach does not explicitly build a 
 phylogenetic network. In Section 8 we briefly discuss the efforts for developing methods for phylogenetic network inference that took place in parallel in the population genetics community (they are often referred to 
 as ``admixture graphs" in the population genetics literature). In Section 9, we summarize the available software for phylogenetic network inference, discuss the data that these methods use, and then list some of the 
 limitations of these methods in practice. We conclude with final remarks and directions for future research in Section 10.

\section{Background for Non-biologists}
In this section we define the biological terminology used throughout the chapter so that it is accessible for non-biologists. 
We also provide a brief review of phylogenetic trees and their likelihood, which is the basis for maximum 
likelihood and Bayesian inference of phylogenetic trees from molecular sequence data. Excellent books that cover mathematical 
and computational aspects of phylogenetic inference include \cite{felsenstein2004inferring,semple-book,gascuel2005mathematics,tandybook,steel2016phylogeny}. 
\subsection{Terminology}
\begin{figure}[!ht]
  \begin{center}
  \includegraphics[width=4.5in]{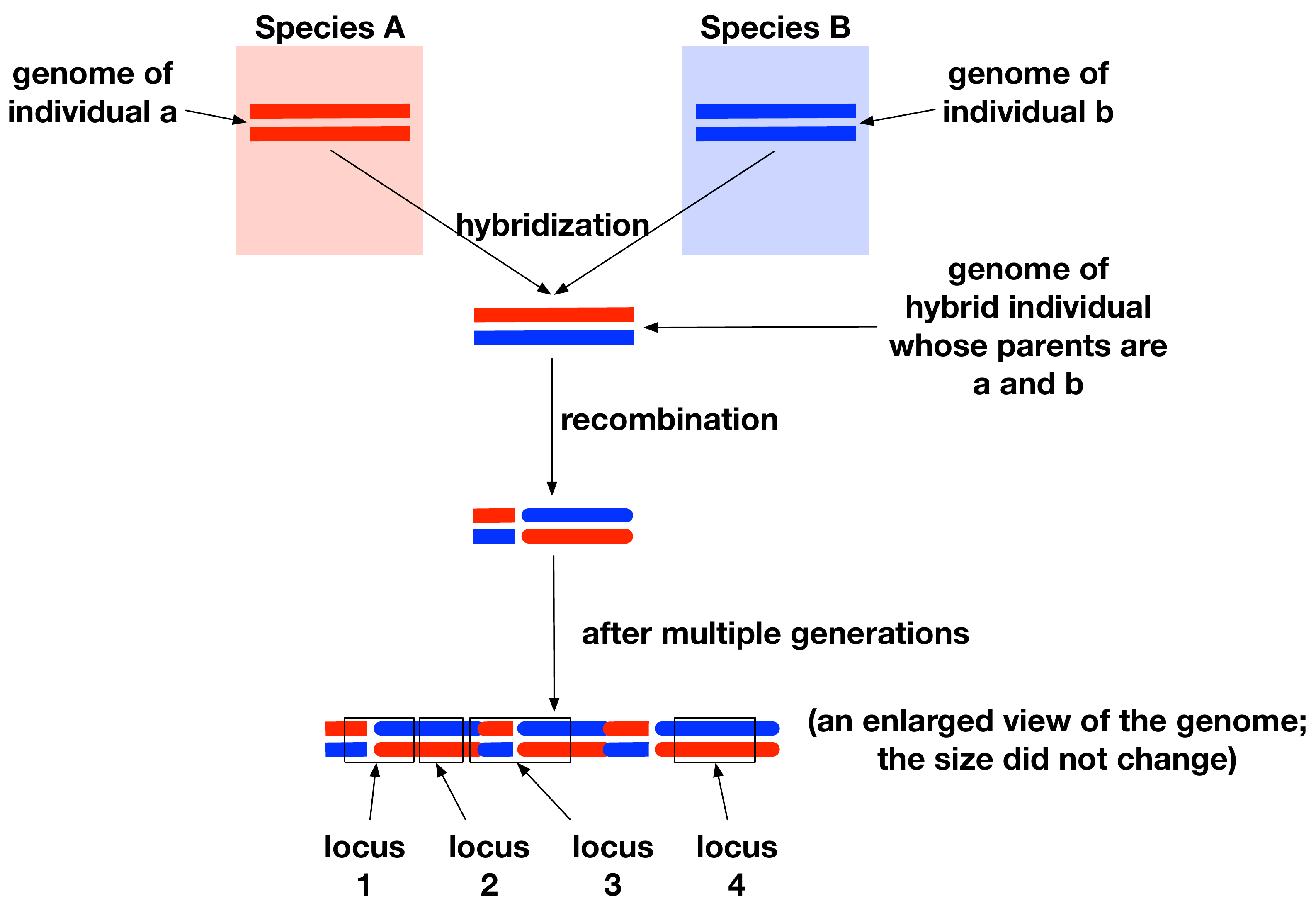}
\caption{{\bf Hybridization, recombination, and the generation of a mosaic genome.} Diploid individual a from species A and individual b from species B mate, resulting in 
a diploid hybrid individual with one copy of its genome inherited from parent a and the other copy inherited from parent b. A recombination event results in the ``swapping" of entire regions 
between two copies of the genome. After multiple generations in which more recombination happens, the genome becomes a mosaic. Walking across the genome 
from left to right, the color switches back and forth between red and blue, where switches happen at recombination breakpoints. Shown are four different loci. Loci 1 and 3 are not appropriate 
for tree inference since they span recombination breakpoints and, thus, include segments that have different evolutionary histories. Loci 2 and 4 are the ``ideal" loci for analyses by methods 
described in this chapter.\label{fig:term}}
\end{center}
\end{figure}

 As we mentioned above, hybridization is reproduction between two members of genetically distinct populations, or species (Fig. \ref{fig:term}). 
 Diploid species (e.g., humans) have two copies of each genome. Aside from a few unusual organisms such as parthenogenic species, one copy will be maternal in origin and the other paternal. When the hybrid individual (or $F_1$) is also diploid this process is called homoploid hybridization. 

While each of the two copies of the genome in the hybrid individual traces its evolution back to precisely one of the two parents, this picture becomes 
much more complex after several rounds of recombination. Recombination is the swapping of a stretch of DNA between the two copies 
of the genome. Mathematically, if the two copies of the genome are given by strings $u$ and $v$ (for DNA, the alphabet for the strings is $\{A,C,T,G\}$), then 
recombination results in two strings $u'=u_1u_2u_3$ and $v'=v_1v_2v_3$, where $u_1$, $u_2$, $u_3$, $v_1$, $v_2$, and $v_3$ are all strings over the same 
alphabet, and $u_1v_2u_3=u$ and $v_1u_2v_3=v$; that is, substrings $u_2$ and $v_2$ were swapped. Observe that when this happens, $u_1$ and $u_3$ in 
the copy $u'$ are inherited from one parent, and $v_2$, also in the copy $u'$, is inherited from a different parent. A similar scenario happens in copy $v'$ of 
the diploid genome. 

This picture gets further complicated due to backcrossing, which is the mating between the hybrid individual, or one of its descendants, with an individual in one 
of the parental species. For example, consider a scenario in which descendants of the hybrid individual in Fig. \ref{fig:term} repeatedly mate with individuals 
from species A. After several generations, it is expected that the genomes of the hybrid individuals become more similar to the genomes of individuals in species 
A, and less similar to the genomes of individuals in species B (using the illustration of Fig. \ref{fig:term}, the two copies of the genome would have much more red 
in them than blue). 

Most models and methods for phylogenetic inference assume the two copies of a diploid genome are known separately and often only one of them is used to represent the corresponding 
individual. However, it is important to note that knowing the two copies separately is not a trivial task. Sequencing technologies produce data on both copies simultaneously, 
and separating them into their constituent copies is a well-studied computational problem known as genome phasing. 

Biologists often focus on certain regions within the genomes for phylogenetic inference. If we consider the genome to be represented as a string $w$ over the alphabet 
$\{A,C,T,G\}$, then a locus is simply a substring of $w$ given by the start and end positions of the substring in $w$. The size of a locus can range anywhere from a single position in the genome to 
a (contiguous) stretch of 1 million or more positions in the genome. As we discussed above, when recombination happens, an individual copy of the genome would have segments 
with different ancestries (the blue and red regions in Fig. \ref{fig:term}). A major assumption underlying phylogenetic tree inference is that the sequence data of a locus used for inference 
has evolved down a single tree. Therefore, the more recombination which has occured within a locus over its evolutionary history (limited to the history connecting the species being studied), the
less suitable it will be for phylogenetic inference. Conversely, loci with low recombination rates may be more suitable in terms of avoiding intra-locus recombination, although such loci are more susceptible to linked selection \cite{hahn2008evolution}.

\subsection{Phylogenetic Trees and Their Likelihood}
An unrooted binary phylogenetic tree $T$ on set $\X$ of taxa (e.g., $\X=\{humans,chimp,gorilla\}$) is a binary tree whose leaves are bijectively labeled by the elements of $\X$. 
That is, if $|\X|=n$, then $T$ has $n$ leaf nodes and $n-2$ non-leaf (internal) nodes (each leaf node has degree $1$ and each internal node has degree $3$). A rooted binary phylogenetic tree is a directed binary tree with a single node designated
as the root and all edges are directed away from the root. For $n$ taxa, a rooted binary tree has $n$ leaves and $n-1$ internal nodes (each leaf node has in-degree $1$ and out-degree $0$;
each internal node except for the root has in-degree $1$ and out-degree $2$; the root has in-degree $0$ and out-degree $2$). 

\begin{figure}[!ht]
  \begin{center}
  \includegraphics[width=4.5in]{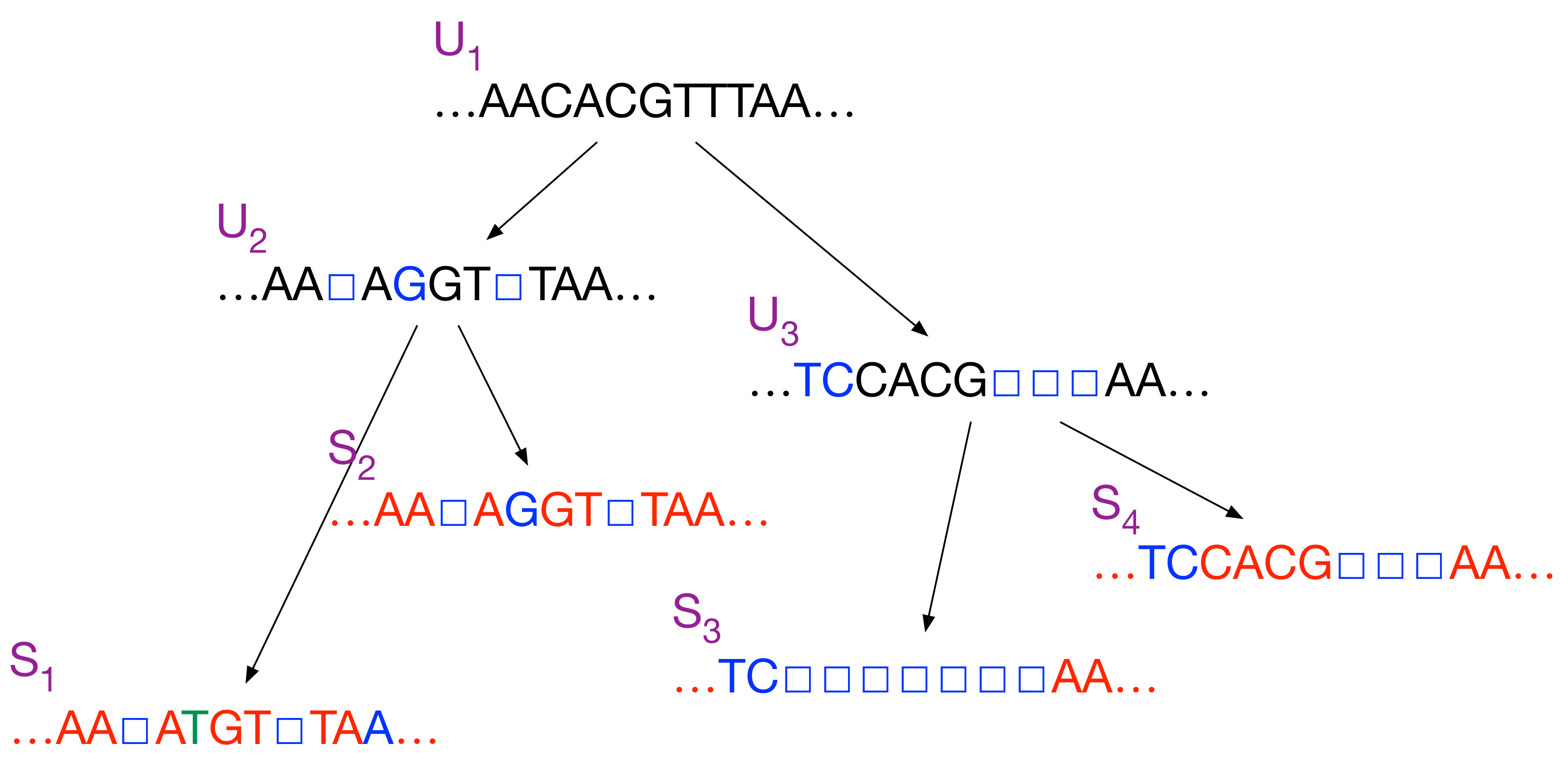}
\caption{{\bf Sequence evolution on a tree.} At the top is the ancestral sequence for a certain locus in the genome of an individual. Through cell division and DNA replication, this 
sequence is inherited from parent to children. However, mutations could alter the inherited sequences. Boxes indicate letters that were deleted due to mutation. Letters in blue 
 indicate substitutions (a mutation that alters the state of the nucleotide). The letter in green has mutated more than once during its evolutionary history. With respect to sequence 
 $U_1$, sequence $U_2$ has two deletions at the 3rd and 8th positions, and a substitution (C to G) at the 5th position. With respect to sequence $U_1$, sequence 
 $S_3$ has 7 deletions at positions 3--9, and substitutions at the first two positions (A to T, and A to C). Sequences $S_1$, $S_2$, $S_3$, and $S_4$, with the boxes and colors 
 unknown, are often the data for 
 phylogenetic inference. That is, the four sequences used as data here are: AAATGTTAA, AAAGGTTAA, TCAA, and TCCACGAA. \label{fig:seqevol}}
\end{center}
\end{figure}

Modern methods for phylogenetic tree inference make use of molecular sequence data, such as DNA sequences, obtained from individuals within the species of interest. 
The sequences are assumed to have evolved from a common ancestral sequence (we say the sequences are homologous) according to a model of evolution that specifies the rates 
at which the various mutational events could occur (Fig. \ref{fig:seqevol}). 

For example, to infer a phylogenetic tree $T$ on set $\X=\{X_1,\ldots,X_n\}$ of taxa, the sequence $S_1$ of a certain locus is obtained from the genome of an individual in species 
$X_1$, the sequence $S_2$ of a certain locus is obtained from the genome of an individual in species $X_2$, and so on until $n$ sequences $S_1,\ldots,S_n$ are obtained. To 
perform phylogenetic tree inference, the $n$ sequences must satisfy two important conditions (see Fig. \ref{fig:seqevol}): 
\begin{itemize}
\item {\em The sequences are homologous:} The obtained sequences must have evolved down a single tree from a single sequence in an individual in an ancestral species. Two sequences are homologous if they evolved from a common ancestor, including in the presence of events such as duplication. Two homologous sequences are orthologs if they evolved from a common 
ancestor solely by means of DNA replication and speciation events. Two homologous sequences are paralogs if their common ancestor had duplicated to give rise to the two sequences.
\item {\em The sequences are aligned:} While the obtained homologous  ``raw" sequences might be of different lengths due to events such as insertions and deletions, the sequences must be made 
to be the same length before phylogenetic inference is conducted so that positional homology is established. Intuitively, positional homology is the (evolutionary) correspondence among sites 
across the $n$ sequences. That is, the sequences must be made of the same length so that the $i$-th site in all of them had evolved from a single site in the sequence that is ancestral to 
all of them. 
\end{itemize}
 \begin{figure}[!ht]
  \begin{center}
  \includegraphics[width=4.5in]{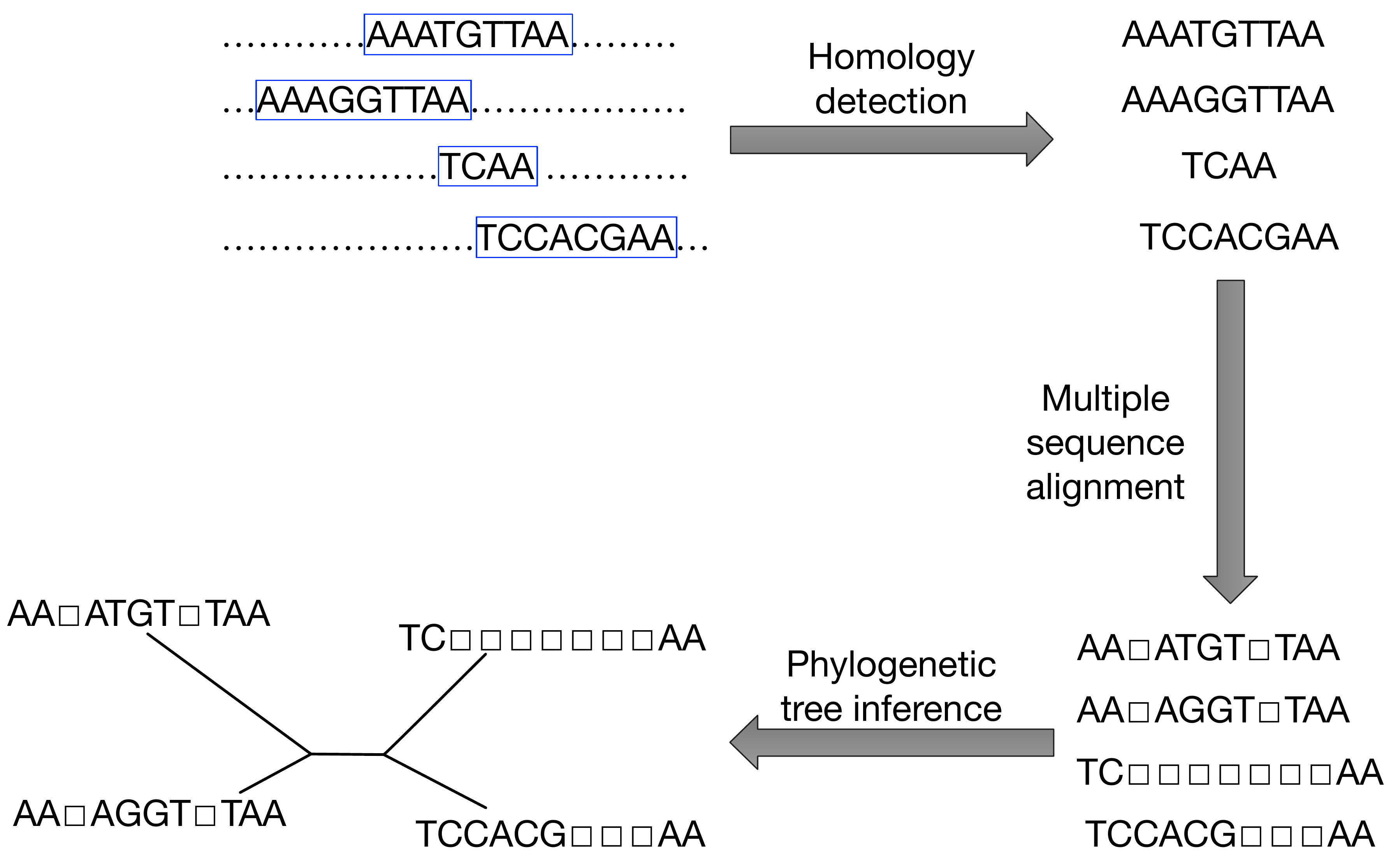}
\caption{{\bf From homologous sequences to a phylogenetic tree.} Identifying homologous sequences across genomes is the first step towards a phylogenetic analysis. The homologous 
 sequences, once identified, are not necessarily of the same length, due to insertions and deletions. Multiple sequence alignment is performed on the homologous sequences and the result 
 is sequences of the same length where boxes indicate deleted nucleotides. Finally, a phylogenetic tree is constructed on the aligned sequences.  \label{fig:phylopipeline}}
\end{center}
\end{figure}

Identifying homologous sequences across genomes is not an easy task; see, for example, \cite{nichio2017new} for a recent review of methods for homology detection. Multiple sequence 
alignment is also a hard computational problem, with a wide array of heuristics and computer programs currently available for it; see, for example, \cite{chatzou2015multiple} for a recent review. 

We are now in position to define a basic version of the Phylogeny Inference Problem:
\begin{itemize}
\item[]{\bf Input:} Set $S=\{S_1,\ldots,S_n\}$ of homologous sequences, where sequence $S_i$ is obtained from taxon $X_i$, and the $n$ sequences are aligned. 
\item[]{\bf Output:} A phylogenetic tree $T$ on set $\X$ of taxa such that $T$ is optimal, given the sequences, with respect to some criterion $\Phi$. 
\end{itemize}
The books we cited above give a great survey of the various criterion that $\Phi$ could take, as well as algorithms and heuristics for inferring optimal 
trees under the different criteria. Here, we focus on the main criterion in statistical phylogenetic inference, namely likelihood. We will make two assumptions when defining the likelihood that are (1) sites are identically and independently distributed, and (2) following a DNA replication event, the two resulting 
sequences continue to evolve independently of each other. 

To define the likelihood of a tree $T$, we first assign lengths $\lambda:E(T) \rightarrow {\mathbb R}^+$ to its branches, so that $\lambda(b)$ is the length of branch 
$b$ in units of expected number of mutations per site per generation. Furthermore, we need a model of sequence evolution ${\cal M}$. Most models of sequence evolution 
are Markov processes where the probability of observing a sequence $S$ at node $u$ depends only on the sequence at $u$'s parent, the length of the 
branch that links $u$ to its parent, and the parameters of the model of sequence evolution. If we denote by $p^{(i)}_{uv}(t)$ the probability that the 
$i$-th nucleotide in the sequence at node $u$ evolves into the $i$-th nucleotide in the sequence at node $v$ over time $t$ (measured in units of expected number of mutations as well), then 
the likelihood of a tree $T$ and its branch lengths $\lambda$ is 
\begin{equation}
\label{eq:fels}
L(T,\lambda|S) = P(S|T,\lambda) = \prod_{i} \left(\sum_R \left(p(root^{(i)}) \cdot \prod_{b=(u,v)\in E(T)} p^{(i)}_{uv}(\lambda_b) \right) \right).
\end{equation}
Here, the outer product is taken over all sites $i$ in the sequences; i.e., if each of the $n$ sequences is of length $m$, then $1 \leq i \leq m$. 
The summation is taken over $R$, which is the set of all possible labelings of the internal nodes of $T$ with sequences of length $m$. Inside the summation,
$p(root^{(i)})$ gives the stationary distribution of the nucleotides at position $i$. The likelihood as given by Eq. \eqref{eq:fels} is computed in polynomial time in 
$m$ and $n$ using Felsenstein's ``pruning" algorithm \cite{Felsenstein1981}. 

Finally, the maximum likelihood estimate for solving the Phylogeny Inference Problem is given by 
$$(T^*,\lambda^*) \gets {\rm argmax}_{(T,\lambda)} L(T,\lambda|S).$$
 Computing the maximum likelihood estimate from a set $S$ of sequences is NP-hard \cite{Chor2005,rochnphard}. However, much progress has been made in terms of developing 
 heuristics that scale up to thousands of taxa while achieving high accuracy, e.g., \cite{raxml}.

\section{From Humble Beginnings: Smallest Displaying Networks}
 Early work and, still, much effort in the community has focused on inferring the topology of a phylogenetic network from a set of gene tree 
 topologies estimated for the individual loci in a data set. In this section, we discuss parsimony approaches to inferring phylogenetic network 
 topologies from sets of gene trees. 
\subsection{The Topology of a Phylogenetic Network}
\begin{figure}[!ht]
\centerline{\includegraphics[width=0.5\textwidth]{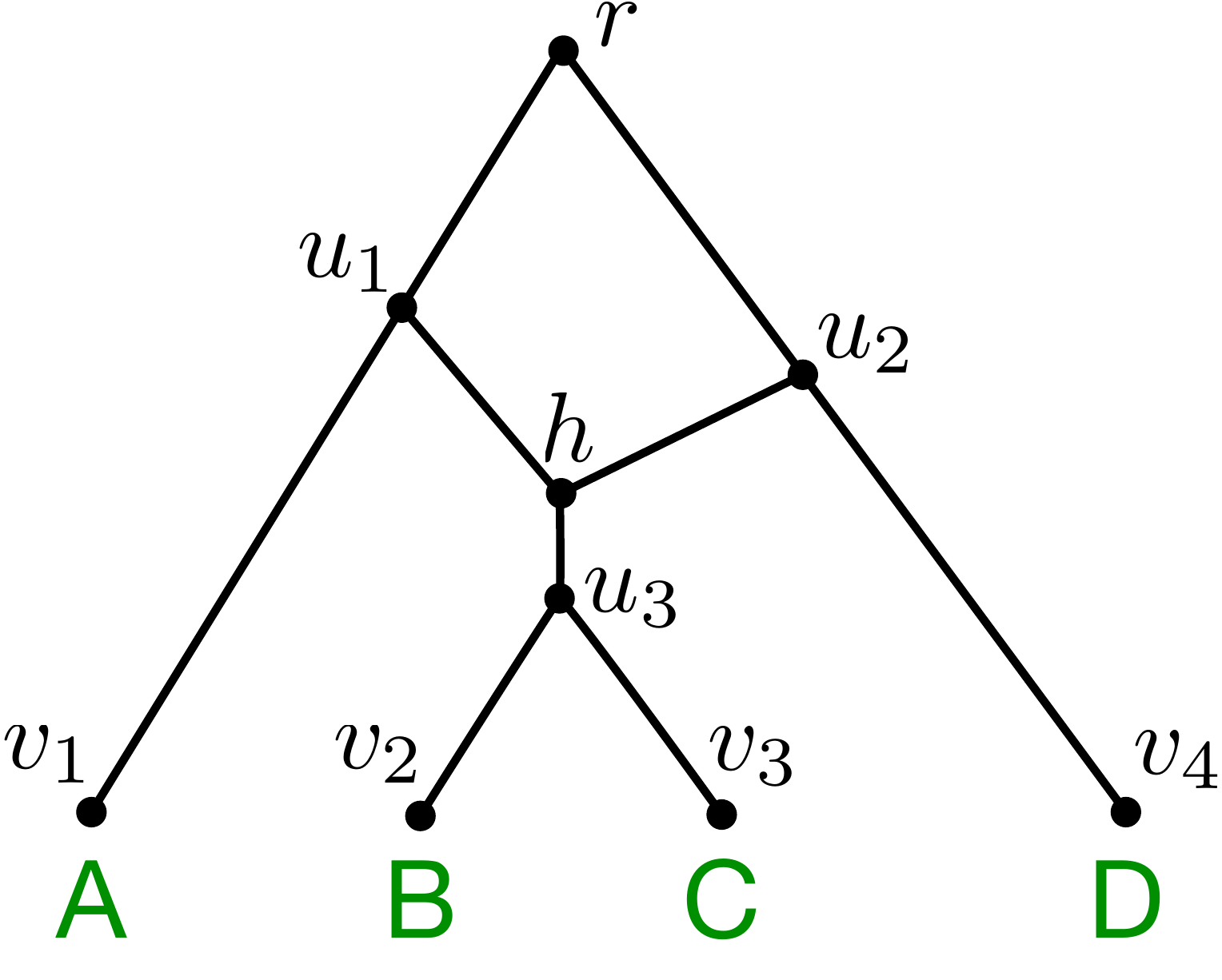}}
\caption{{\bf An example of a phylogenetic network $\Psi$ with a single reticulation event.} This network is made up of leaf nodes $V_L = \{v_1,v_2,v_3,v_4\}$, internal tree nodes $V_T=\{u_1,u_2,u_3\}$, reticulation nodes $V_N=\{h\}$, and the root $r$. The nodes are connected by branches belonging to the set of phylogenetic network edges $E(\Psi)$. The branches are: 
 $(r,u_1)$, $(r,u_2)$, $(u_1,h)$, $(u_2,h)$, $(h,u_3)$, $(u_1,v_1)$, $(u_2,v_4)$, $(u_3,v_2)$, and $(u_3,v_3)$. The leaves are labeled by set $\X=\{A,B,C,D\}$ of taxa: $\ell(v_1)=A$, $\ell(v_2)=B$, $\ell(v_3)=C$, and $\ell(v_4)=D$. \label{fig:phylogeneticNetwork}}
\end{figure}

 As discussed above, a reticulate, i.e., non-treelike, evolutionary history that arises in the presence of processes such as hybridization and horizontal gene transfer is best represented by a phylogenetic network.
 \begin{definition}
 \label{def:pn}
  A {\em phylogenetic $\X$-network} (Fig.~\ref{fig:phylogeneticNetwork}), or $\X$-network for short, $\Psi$ is a rooted, directed, acyclic graph (rDAG) with set of nodes $V(\Psi)=\{r\} \cup V_L \cup V_T \cup V_N $, where
        \begin{itemize}
        \item $indeg(r)=0$ ($r$ is the {\em root} of $\Psi$);
        \item $\forall{v \in V_L}$, $indeg(v)=1$ and $outdeg(v)=0$ ($V_L$ are the {\em external tree nodes}, or {\em leaves}, of $\Psi$);
        \item $\forall{v \in V_T}$, $indeg(v)=1$ and $outdeg(v) \geq 2$ ($V_T$ are the {\em internal tree nodes} of $\Psi$); and, 
        \item $\forall{v \in V_N}$, $indeg(v) = 2$ and $outdeg(v) = 1$ ($V_N$ are the {\em reticulation nodes}
         of $\Psi$).
        \end{itemize}
 For {\em binary} phylogenetic networks, the out-degree of the root and every internal tree node is $2$. 
 The network's set of edges, denoted by $E(\Psi) \subseteq V \times V$ is bipartitioned into {\em reticulation edges}, whose heads are reticulation nodes, and {\em tree edges}, whose heads are tree nodes (internal or external). Finally, the leaves of $\Psi$ are bijectively labeled by  the {\em leaf-labeling} function
  $\ell:V_L \rightarrow \X$.
  \end{definition}
\subsection{Inferring Smallest Displaying Networks}
 Early work on phylogenetic networks focused on the problem of identifying a network with the fewest number of reticulation nodes 
 that summarizes all gene trees in the input. More formally, let $\Psi$ be a phylogenetic network. We say that $\Psi$ {\em displays} 
 phylogenetic tree $t$ if $t$ can be obtained from $\Psi$ by repeatedly applying the following operations until they are not applicable:
 \begin{enumerate}
 \item For a reticulation node $h$ with two incoming edges $e_1=(u_1,h)$ and $e_2=(u_2,h)$, remove one of the two edges. 
 \item For a node $u$ with a single parent $v$ and a single child $w$, remove the two edges $(v,u)$ and $(u,w)$, and add edge $(v,w)$. 
 \end{enumerate}
 The set of all trees displayed by the phylogenetic network is 
 $${\cal T}(\Psi)=\{t: \; \Psi \; {\rm displays} \; t\}.$$ 
 For example, for the phylogenetic network $\Psi$ of Fig. \ref{fig:phylogeneticNetwork}, we have ${\cal T}(\Psi)=\{T_1,T_2\}$, where 
 $T_1=((A,(B,C)),D)$ and $T_2=(A,((B,C),D))$.
  
 Using this definition, the earliest phylogenetic network inference problem was defined as follows:
 \begin{itemize}
 \item[]{\bf Input:} A set ${\cal G}=\{g_1,g_2,\ldots,g_m\}$ of gene trees, where $g_i$ is a gene tree for locus $i$. 
 \item[]{\bf Output:} A phylogenetic network $\Psi$ with the smallest number of reticulation nodes such that ${\cal G} \subseteq {\cal T}(\Psi)$. 
 \end{itemize}
 This problem is NP-hard \cite{Wang2001} and methods were developed for solving it and variations thereof, some of which are heuristics \cite{van2010phylogenetic,wu2010close,ParkNakhleh12,wu2013algorithm}. Furthermore, the view of a phylogenetic network 
 in terms of the set of trees it displays was used for pursuing other questions in this domain. For example, the topological difference between 
 two networks could be quantified in terms of the topological differences among their displayed trees \cite{nakhleh2002towards}. The parsimony and likelihood criteria 
 were extended to the case of phylogenetic networks based on the assumption that each site (or, locus) has evolved down one of the trees 
 displayed by the network \cite{NakhlehJin05,JinNakhleh-bioinfo,JinNakhleh-eccb,JinNakhleh-isbra,JinNakhleh-mbe,JinNakhleh-tcbb}. The concepts of character compatibility and perfect phylogeny were also extended to phylogenetic networks based 
 on the notion of displayed trees \cite{Wang2001,nakhleh2005perfect,KanjNakhleh08a}. Furthermore, questions related to distinguishability of phylogenetic networks based on their displayed trees 
 have been pursued \cite{KanjNakhleh08b} and relationships between networks and trees have been established in terms of this definition \cite{francis2015phylogenetic,zhang2016tree}.  

 However, the computational complexity of this 
 problem notwithstanding, the problem formulation could be deficient with respect to practical applications. For one thing, solving the aforementioned 
 problem only yields the topology of a phylogenetic network, but no other parameters. In practice, biologists would be interested in divergence times, 
 population parameters, and some quantification of the amount of introgression in the genomes. These quantities are not recoverable under the given 
 formulation. Moreover, for the biologist seeking to 
 analyze her data with respect to hypotheses of reticulate evolutionary events, solving the aforementioned problem could result in misleading 
 evolutionary scenarios for at least three reasons. First, the smallest number of reticulations required in a phylogenetic network to display all trees in the 
 input could be arbitrarily far from the true (unknown) number of reticulations. One reason for this phenomenon is the occurrence of reticulations between 
 sister taxa, which would not be detectable from gene tree topologies alone. Second, a smallest set of reticulations could be very different from the actual 
 reticulation events that took place. Third, and probably most importantly, some or even all of gene tree incongruence in an empirical data set could have 
 nothing to do with reticulation. For example, hidden paralogy and/or incomplete lineage sorting could also give rise to incongruence in gene trees. When such 
 phenomena are at play, seeking a smallest phylogenetic network that displays all the trees in the input is the wrong approach and might result in an overly complex 
 network that is very far from the true evolutionary history. To address all these issues, the community has shifted its attention in the last decade toward 
 statistical approaches that view phylogenetic networks in terms of a probability distribution on gene trees that could encompass a variety of evolutionary 
 processes, including incomplete lineage sorting. 
 
 \subsection{Phylogenetic Networks as Summaries of Trees}
\begin{figure}[!ht]
 \begin{center}
 \includegraphics[width=4in]{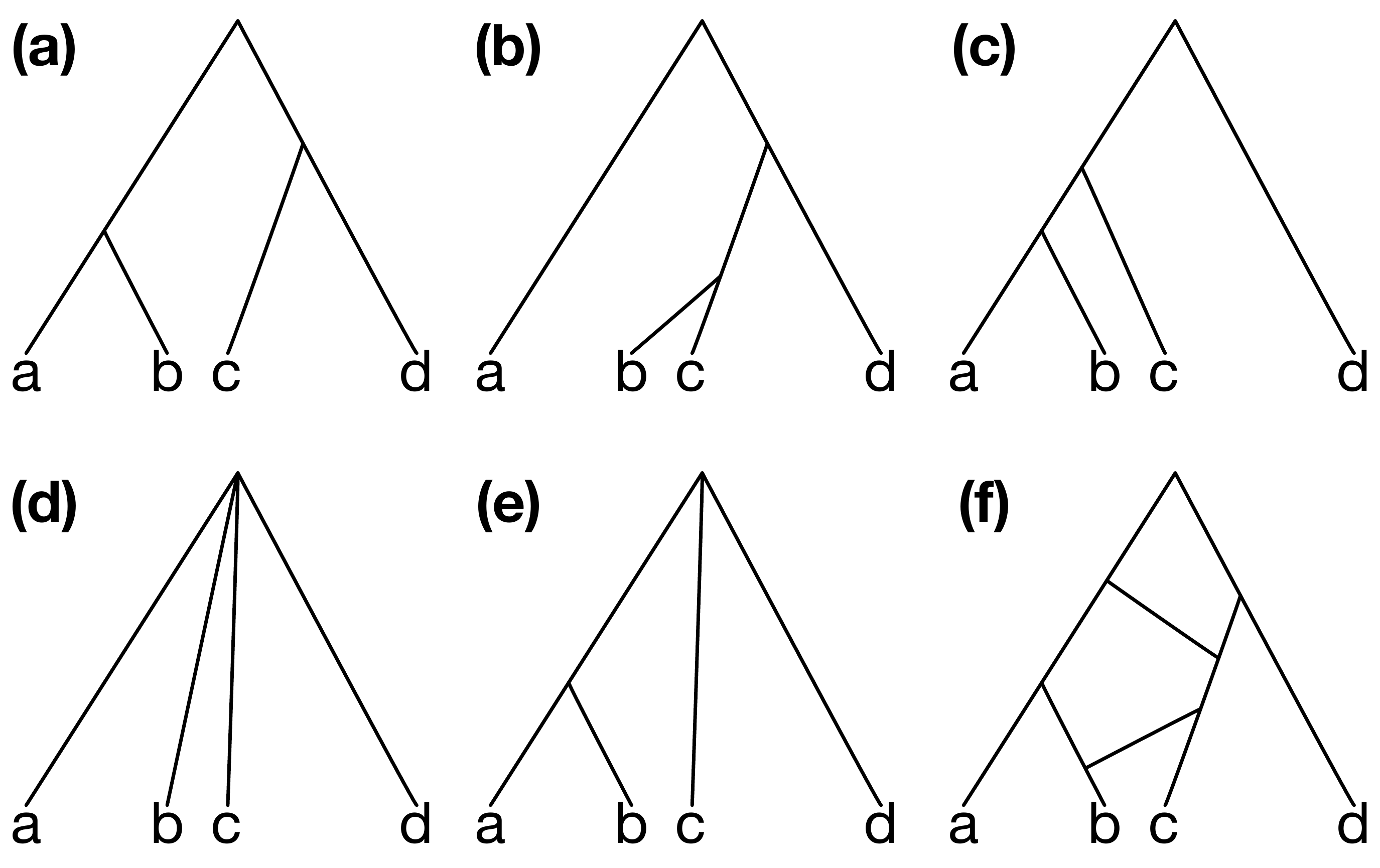}
 \caption{{\bf Consensus trees and phylogenetic networks as two contrasting summary methods.} (a)-(c) Three (input) gene trees whose summary is sought. 
 (d) The strict consensus of the input trees. (e) The 70\% majority-rule consensus of the input trees. (f) A smallest phylogenetic network that displays all three trees i
 in the input. The strict consensus could be resolved to yield 15 different binary trees, only three of which are in the input. The majority-rule consensus tree could be 
 resolved to yield three possible trees, two of which are the trees in (a) and (d), but the third, which is (((a,b),d),c), is not in the input. Furthermore, the tree in panel (b) 
 is not included in the summary provided by the majority-rule consensus. The phylogenetic network displays four trees, three of which are the input trees, and the 
 fourth is ((a,(b,c)),d), which is not in the input.\label{fig:summaryphylo}}
 \end{center}
 \end{figure}

 Before we turn our attention to these statistical approaches, it is worth contrasting smallest phylogenetic networks that display all trees in the input to the 
 concept of consensus trees. In the domain of phylogenetic trees, consensus trees have played an important role in compactly summarizing sets of 
 trees. For example, the strict consensus tree contains only the clusters that are present in the input set of trees, and nothing else. The majority-rule 
 consensus tree contains only the clusters that appear in at least 50\% of the input trees. When there is incongruence in the set of trees, these consensus trees 
 are most often non-binary trees (contain ``soft polytomies") such that each of the input trees can be obtained as a binary resolution of the consensus tree. Notice 
 that while the consensus tree could be resolved to yield each tree in the input, there is no guarantee in most cases that it cannot also be resolved to generate trees 
 that are not in the input. Smallest phylogenetic networks that display all trees in the input could also be viewed as summaries of the trees, but instead of removing 
 clusters that are not present in some trees in the input, they display all clusters that are present in all trees in the input. Similarly to consensus trees, a smallest 
 phylogenetic network could also display trees not in the input (which is the reason why we use $\subseteq$, rather than $=$, in the problem formulation above). 
 These issues are illustrated in Fig. \ref{fig:summaryphylo}. 
 
 As discussed above, ILS is another process that could cause gene trees to be incongruent with each other and complicates the inference of 
 phylogenetic networks since incongruence due to ILS should not induce additional reticulation nodes. Before we move on to discuss statistical 
 approaches that account for ILS in a principled probabilistic manner under the coalescent, we describe an extension of the {\em minimizing deep coalescences}, 
 or, MDC, criterion \cite{Maddison97,MaddisonAndKnowles06,ThanNakhleh-PLoSCB09}, to phylogenetic networks, which was devised in \cite{YuEtAl13}.
 
 \subsection{A Step Towards More Complexity: Minimizing Deep Coalescences}
 Let $\Psi$ be a phylogenetic network and consider node $u \in V(\Psi)$. We denote by $B_u \subseteq V(\Psi)$ the set of nodes in $\Psi$ that are below node $u$ (that is, the set of nodes that are reachable from the root of $\Psi$ via at least one path that goes through node $u$). 
 \begin{definition}
 A coalescent history of a gene tree $g$ and 
a species (phylogenetic) network $\Psi$ as a function $h:V(g) \rightarrow V(\Psi)$ such that the following conditions hold:
\begin{itemize}
\item  if $w$ is a leaf in $g$, then $h(w)$ is
the leaf in $\Psi$ with the same label (in the case of multiple alleles, $h(w)$ is the leaf in $\Psi$ with the label of the species from which the 
allele labeling leaf $w$ in $g$ is sampled); and, 
\item if $w$ is a node in $g_v$, then $h(w)$
is a node in $B_{h(v)}$. 
 \end{itemize}
 \end{definition}
Given a phylogenetic network $\Psi$ and a gene tree $g$, we denote by $H_{\Psi}(g)$ the set of all coalescent histories of gene tree $g$ within the branches of 
phylogenetic network $\Psi$. 

Given a coalescent history $h$, the {\it number of extra lineages} arising from $h$ on a branch $b=(u,v)$ in 
 phylogenetic network $\Psi$ is the number of gene tree lineages exiting branch $b$ from below node $u$ toward the root, minus one. 
 Finally, $XL(\Psi,h)$ is defined as the sum of the numbers of extra lineages arising from $h$ on 
 all branches $b \in E(\Psi)$. 

Using coalescent histories, the minimum number of extra lineages required to 
reconcile gene tree $g$ within the branches of $\Psi$, denoted by $XL(\Psi,g)$ is given by 
\begin{equation}
\label{eq:xlUsingCH}
XL(\Psi,g)=\min_{h \in H_{\Psi}(g)}XL(\Psi,h). 
\end{equation}
Under the MDC (minimizing deep coalescence) criterion, the optimal coalescent history refers to the one that results in the fewest number of extra lineages  
 \cite{Maddison97,ThanNakhleh-PLoSCB09}, and thus,
\begin{equation}
\label{eq:xlNetwork}
XL(\Psi, g) = \sum_{e \in E(\Psi)}[k_e(g)-1]
\end{equation}
where $k_e(g)$ is the number of extra lineages on edge $e$ of $\Psi$ in the optimal coalescent history of gene tree $g$. 

 \begin{figure}[!ht]
 \begin{center}
 \includegraphics[width=3in]{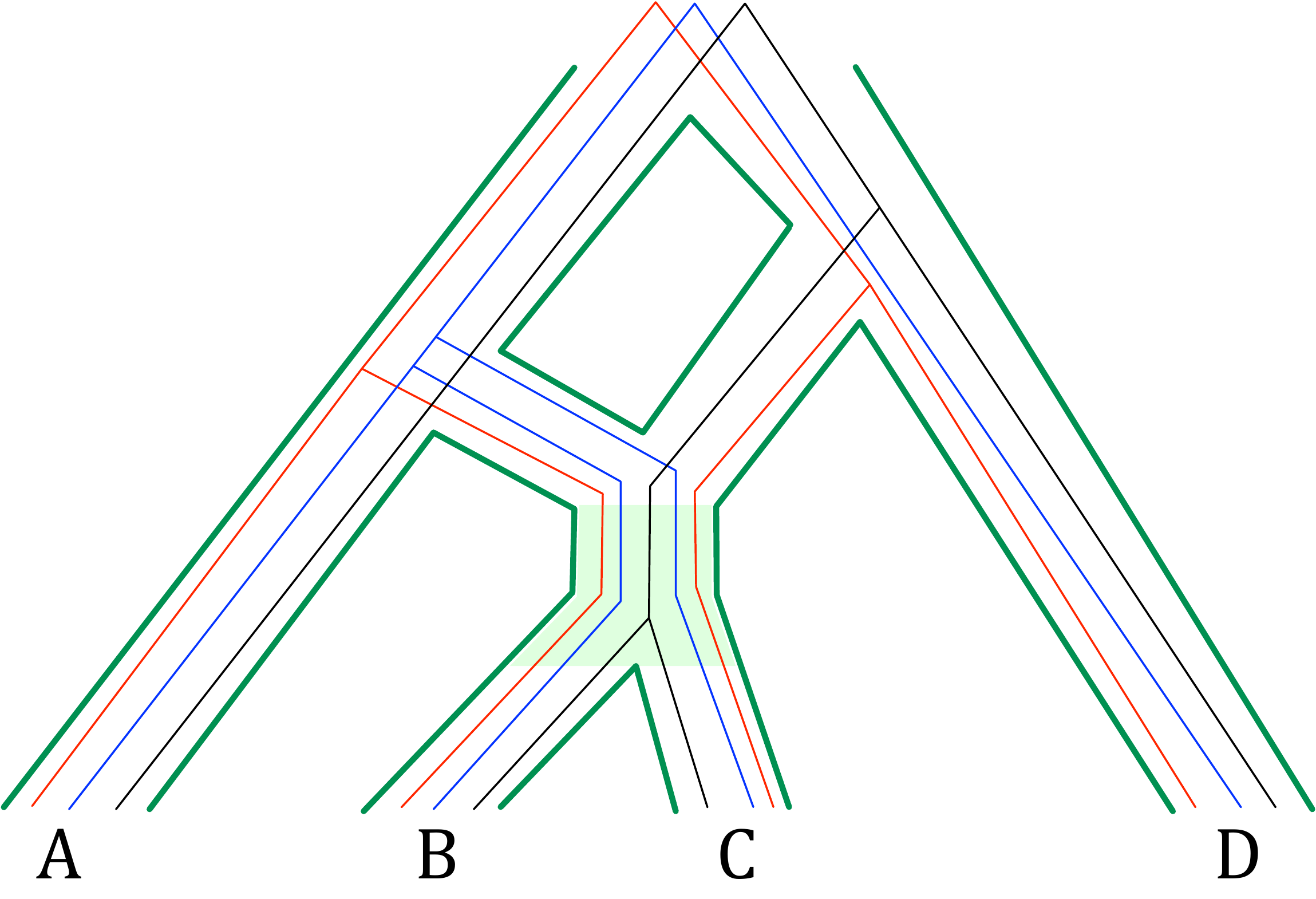}
 \caption{{\bf The MDC criterion on phylogenetic networks.} A phylogenetic network and coalescent histories within its branches of the three gene trees in Fig. \ref{fig:summaryphylo}(a)-(c). The highlighted branch that separates the hybridization event from the MRCA of B and C has two extra lineages arising from the three 
 shown coalescent histories. All other branches have 0 extra lineages. Therefore, the total number of extra lineages in this case is 2. \label{fig:mdc}}
 \end{center}
 \end{figure}

 A connection between extra lineages and the displayed trees of a phylogenetic network is given by the following observation.
 \begin{observation}
If gene tree $g$ is displayed by phylogenetic network $\Psi$, then $XL(\Psi,g)=0$. 
\end{observation}
The implication of this observation is that if one seeks the phylogenetic network that minimizes the number of extra lineages, the problem can 
be trivially solvable by finding an overly complex network that displays every tree in the input. Therefore, inferring a phylogenetic network $\Psi$ from a collection of gene tree topologies $\G$ based on the MDC criterion is more appropriately defined by
$$\hat{\Psi}(m) = \argmin_{\Psi(m)} \left( \sum_{g \in \G} XL(\Psi(m),g) \right),$$
where we write $\Psi(m)$ to denote a phylogenetic network with $m$ reticulation nodes. While the number of reticulations $m$ is unknown and is often a quantity of 
interest, there is a trade-off between the number of reticulation nodes and number of extra lineages in a network: Reticulation edges can be added to reduce the number 
of extra lineages. Observing this reduction in the number of extra lineages could provide a mechanism to determine when to stop adding reticulations to the network 
\cite{YuEtAl13}. 

\section{Phylogenetic Networks: A Generative Model of Molecular Sequence Data}
 In the previous section, we focused on two parsimony formulations for inferring a phylogenetic network from a collection of input gene 
 tree topologies: The first seeks a network with the fewest number of reticulations that displays each of the input gene trees, and the second 
 seeks a network that does not have to display every gene tree in the input, but must minimize the number of ``extra lineages" that could arise 
 within a given number of reticulations. Both formulations result in phylogenetic network topologies alone and make use of only the gene tree 
 topologies. In this section, we introduce the multispecies network coalescent, or MSNC \cite{WenEtAl16a}, as a generative process that extends the 
 popular multispecies coalescent, or MSC \cite{Degnan09}, that is the basis for most multi-locus species tree inference methods. The MSNC allows for 
 the coalescent to operate within the branches of a phylogenetic network by viewing a set of populations---extant and ancestral---glued together 
 by a rooted, directed, acyclic graph structure.

\subsection{Parameterizing the Network's Topology}
 In addition to the topology of a phylogenetic network $\Psi$, as given by Definition \ref{def:pn} above, the nodes and edges are parameterized as follows.
 
 Associated with the nodes are divergence/reticulation times, $\tau: V(\Psi) \rightarrow {\mathbb R}^+$, where $\tau(u)$ is the divergence time associated with 
 tree node $u$ and $\tau(v)$ is the reticulation time associated with reticulation node $v$. All leaf nodes $u$ in the network have $\tau(u)=0$. Furthermore, 
 if $u$ is on a path from the root of the network to a node $v$, then $\tau(u) \geq \tau(v)$. 
 
 Associated with the edges are population mutation rate parameters, $\theta: E(\Psi) \rightarrow {\mathbb R}^+$, where $\theta_b=4N_b \mu$ is the population mutation 
 rate associated with edge $b$, $N_b$ is the effective population size associated with edge $b$, and $\mu$ is the mutation rate per site per generation.
 
Divergence times associated with nodes in the phylogenetic network could be measured in units of years, generations, or coalescent units. 
Branch lengths in gene trees are often given in units of expected number of mutations per site. The following rules are used to convert back and 
forth between these units: 
\begin{itemize}
\item Given divergence time $\tau$ in units of expected number of mutations per site, mutation rate per site per generation $\mu$ and the number of generations per year $g$, $\tau / (\mu g)$ represents divergence times in units of years.
\item Given population size parameter $\theta$ in units of population mutation rate per site, $2\tau / \theta$ represents divergence times in coalescent units.
\end{itemize}

 In addition to the divergence times and population size parameters, the reticulation edges of the network are associated with 
 {\em inheritance probabilities}. For every reticulation node $u \in V_N$, let $left(u)$ and $right(u)$ be the ``left" and ``right" 
 edges incoming into node $u$, respectively (which of the two edges is labeled left and which is labeled right is arbitrary). 
 Let $E_R \subseteq E(\Psi)$ be the set of reticulation edges in the network. 
 The inheritance probabilities are a function $\gamma: E_R \rightarrow [0,1]$ such that for every reticulation node 
 $u \in V_N$, $\gamma(left(u))+\gamma(right(u)) = 1$. In the literature, $\gamma$ is sometimes  described as a vector $\Gamma$.  

\subsection{The Multispecies Network Coalescent and Gene Tree Distributions}
\label{msnc} 

As an orthologous, non-recombining genomic region from a set $\X$ of species evolves within the branches of the species phylogeny of $\X$, the 
 genealogy of this region, also called the {\em gene tree}, can be viewed as a discrete random variable whose values are all possible gene 
 tree topologies on the set of genomic regions. When the gene tree branch lengths are also taken into account, the random variable 
 becomes continuous. Yu \textit{et al.} \cite{YuEtAl12} gave the probability mass function (pmf) for this discrete random variable given the phylogenetic network $\Psi$ and an additional parameter $\Gamma$ that contains the inheritance probabilities 
 associated with reticulation nodes, which we now describe briefly.

\begin{figure}[!ht]
\begin{center}
\includegraphics[width=4.3in]{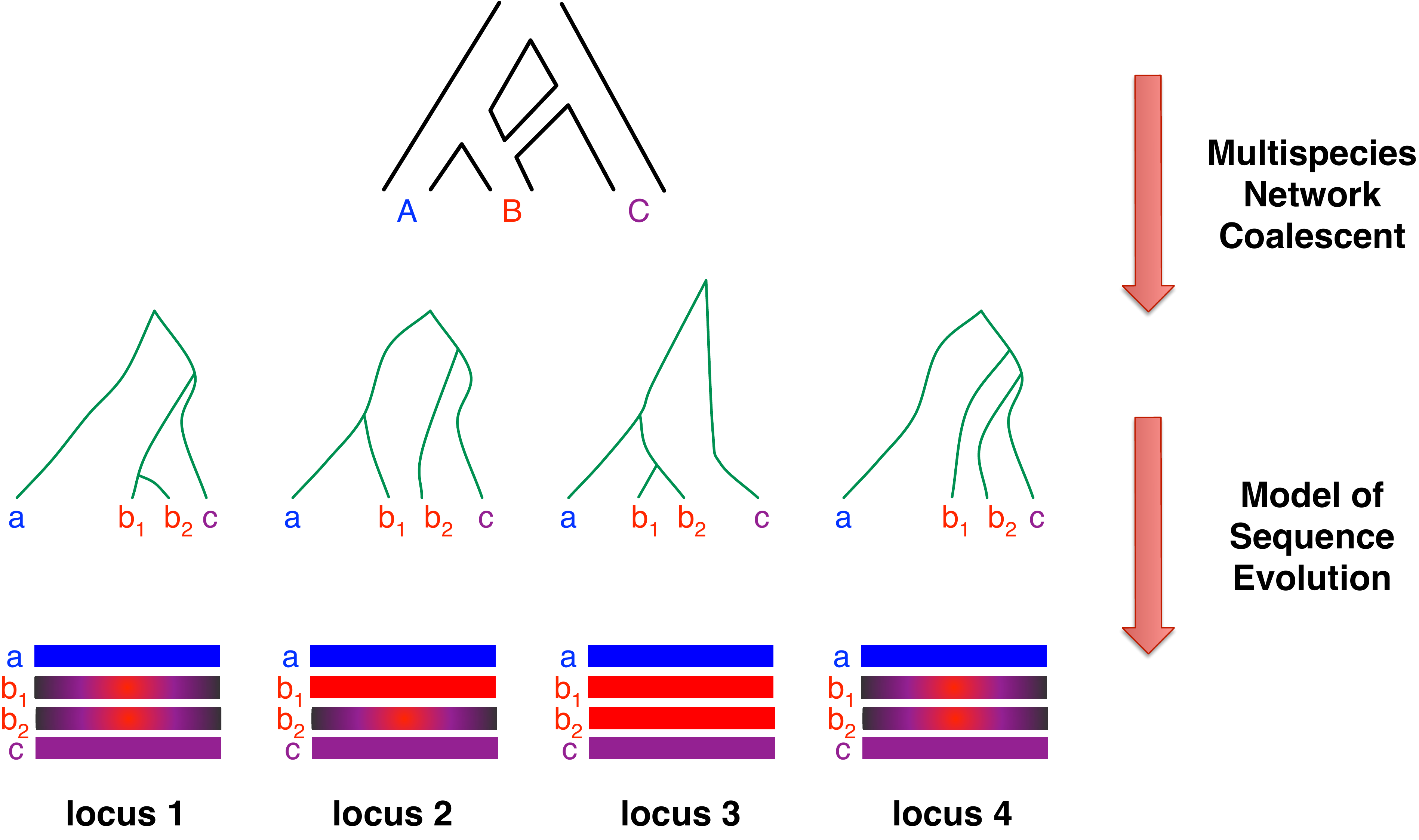}
\caption{{\bf Layers of the multispecies network coalescent model.} A phylogenetic network describes the relationship between species (top). The MSNC describes the distribution of gene trees within the network, in which alleles from the same species can have different topologies and inheritance histories due to reticulation and/or ILS (middle). Some kind of mutation process occurs along the gene trees, resulting in observed differences between alleles in the present, which vary between genes based on their individual trees.} \label{fig:LVM-net}
\end{center}
\end{figure}

  The parameters $\Psi$ and $\Gamma$ specify the {\em multispecies network coalescent}, or MSNC (Fig.~\ref{fig:LVM-net}), and allow for a full 
  derivation of the mass and density functions of gene trees when the evolutionary history of species involves both ILS and 
  reticulation \cite{YuEtAl12,YuEtAl14}. This is a generalization of the {\em multispecies coalescent}, which describes the embedding and distribution of gene trees within a species tree without any reticulate nodes \cite{Degnan09}.

\begin{figure}[!htp]
\centerline{\includegraphics[width=0.8\textwidth]{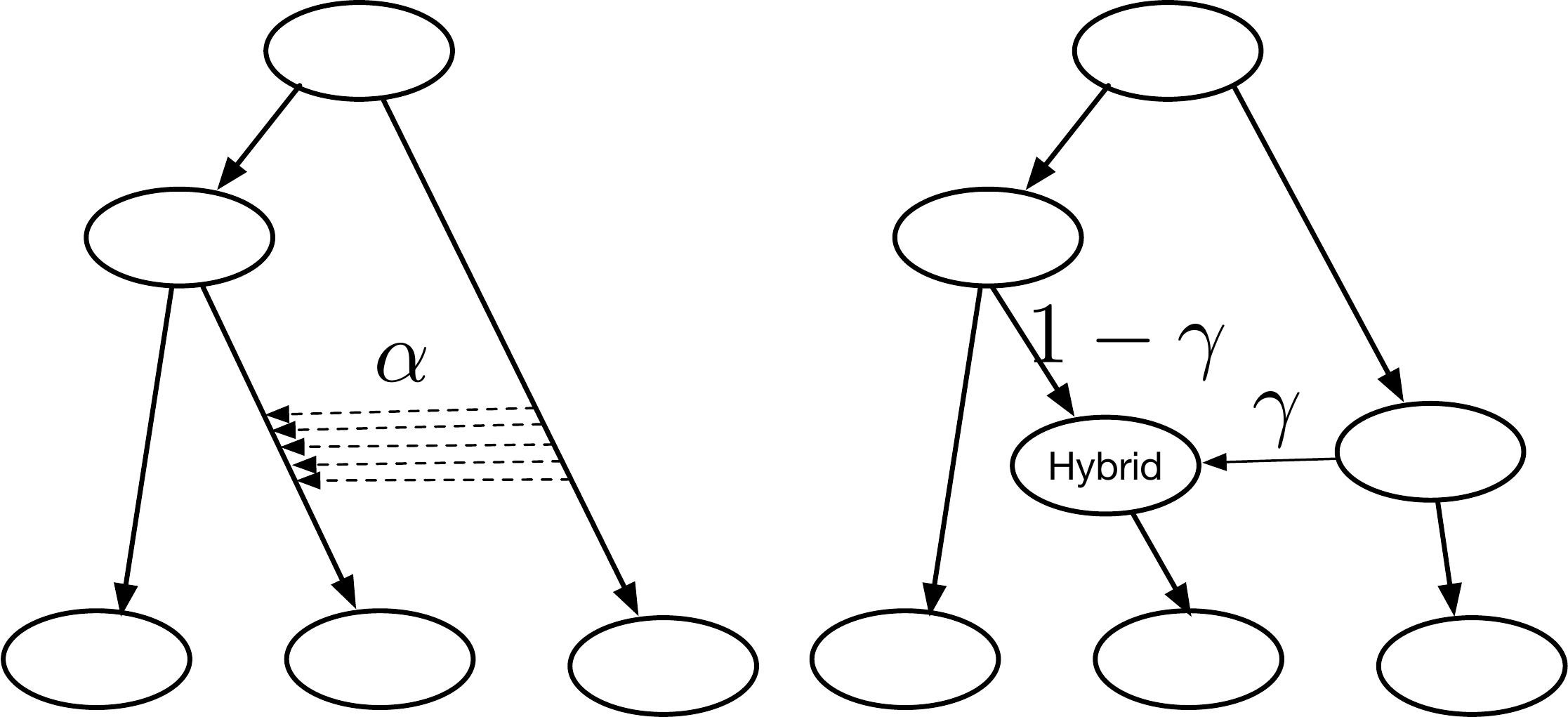}}
\caption{{\bf Reticulation edges as abstractions of gene flow epochs.} (Left) An epoch of gene flow from one population to another with migration rate $\alpha$ per generation. (Right) A phylogenetic network 
with a single reticulation edge that abstracts the gene flow epoch, with inheritance probability $\gamma$. \label{fig:inheritance}}
\end{figure}

It is important to note that a single reticulation edge between two nodes does not mean a single hybridization event. Rather, a reticulation edge abstracts a continuous epoch of repeated gene flow between 
 the two species, as illustrated in Fig.~\ref{fig:inheritance}. 
The two models shown in Fig.~\ref{fig:inheritance} were referred to as the ``gene flow" model (left) and ``intermixture" model (right) of hybridization in \cite{Long91}. 
While the ``gene flow'' model is used by the IM family of methods \cite{doi:10.1093/molbev/msp296} to incorporate admixture, the MSNC adopts the ``intermixture'' model. In this model, the $\gamma$ inheritance probabilities indicate the ratio of genetic materials of a hybrid coming from its two parents. This means that unlinked loci from a hybrid species will have independent evolutionary histories, and will have evolved through the ``left'' or ``right'' parent with some probability $\gamma$ and $1 - \gamma$ respectively. The performance of phylogenetic network inference on data simulated under the gene flow model was demonstrated in \cite{wen2017co}. 

Wen and Nakhleh \cite{wen2017co} derived the density function of the probability of gene trees given a phylogenetic network, with its topology, divergence/migration times, population mutation rates and inheritance probabilities. The divergence/migration times are in units of expected number of mutations per site, and population mutation rates are in units of population mutation rate per site. Based on the MSNC, and by integrating out all possible gene trees, Zhu {\em et al.} \cite{zhu2017bayesian} developed an algorithm to compute the probability of a bi-allelic genetic marker given a phylogenetic network.

\section{Maximum Likelihood Inference of Phylogenetic Networks}
Phylogenetic networks are more complicated than a tree with some reticulation edges. The gene tree topology with highest mass probability may not be one of the backbone trees of the network with 4 or more taxa \cite{ZhuYuNakhleh16}. Also, not all networks can be obtained by simply adding edges between the original edges of a tree \cite{francis2015phylogenetic}. 

Therefore inferring phylogenetic networks is not a trivial extension of methods to infer species trees. Most phylogenetic network methods \cite{YuEtAl13,YuEtAl14,WenEtAl16a,wen2017co,zhang2017bayesian,zhu2017bayesian,Zhu289207,YuNakhleh15b} sample from whole-network space rather than simply adding reticulations to a backbone tree. As such methods walk the space from one phylogenetic network to another, the point estimate or posterior distribution of networks is not tree-based, does not return or imply a backbone tree, and can include networks which cannot be described by merely adding reticulate branches between tree branches.



\subsection{Inference}

Sequential inference was initially developed to estimate species trees under the multispecies coalescent \cite{doi:10.1093/bioinformatics/btu462,Liu2010}, and in recent years has been extended to species networks \cite{YuRisticNakhleh13,YuEtAl14,YuNakhleh15b,solis2016inferring,WenEtAl16a}. These methods follow a two-step approach, where the first step is to estimate gene trees from multiple sequence alignments. The second step is to estimate a species tree or network from the distribution of estimated gene tree topologies.

As described above and illustrated in Fig. \ref{fig:LVM-net}, two key requirements have facilitated the development of several methods for phylogenetic inference from multi-locus data, including those that follow the 
two-step approach. 
 One key requirement for current methods is that the segments are the result of speciation and not gene duplication, that is, sequences from different individuals and species are orthologs and not paralogs. Meeting this requirement ensures that the nodes in each gene tree represent coalescent events and can be fit to a coalescent model within each species network branch. A model which accounts for gene duplication and loss in addition to coalescent events has been developed to reconcile gene family trees with a fixed species tree \cite{Rasmussen01042012,10.1007/978-3-319-59575-7_18}. The most recent implementation of this model can also use it to estimate the species tree \cite{du2018}, but this model has yet to be extended to work with species networks.

A second key requirement is that the evolutionary history of the locus can be accurately modeled using a single tree. Recombination or gene fusion should not have occurred within a locus, otherwise a gene network would be required to model that locus, breaking the MSNC model of gene trees within a species network \cite{YuThan11}. Because of this requirement, multi-locus methods should be used with short contiguous sequences. The results of a previous study on mammal phylogenetics suggest that individual exons are an appropriate target sequence \cite{doi:10.1093/sysbio/syw082}.

Under these two key requirements, each gene tree is considered to be a valid and independent sample from an underlying distribution of gene trees conditional on some unobserved species network. Of course this assumption in sequential inference is violated as the gene trees are only estimates. Particular methods may be more or less sensitive to gene tree estimation errors. For species trees, methods which infer unrooted species trees (e.g. ASTRAL \cite{doi:10.1093/bioinformatics/btu462}) appear to be more robust relative to methods which infer rooted species trees (e.g. MP-EST \cite{Liu2010}). This is because unrooted methods take unrooted gene trees as input and do not rely on correct rooting of the gene trees \cite{SIMMONS201598}.

An estimate $\hat{g}$ of the true gene tree is typically made using phylogenetic likelihood (see Section 2). 
%
%
%
%
 The phylogenetic likelihood of the sequence alignments can be combined sequentially or simultaneously with the MSNC probability densities of the gene trees to estimate a species network from sequence data.

Given gene trees where each node represents a coalescent event, the probability densities and masses of those gene trees given a species network can be calculated \cite{YuEtAl14}. This can be based on the topologies and node heights of the gene trees, or based on the topologies alone (see Section~\ref{msnc}). 

%
%
%
%
%

%
%

Maximum likelihood (ML) methods seek a phylogenetic network (along with its parameters) that maximizes some likelihood function. In a coalescent context, these methods search for the species network which maximizes the likelihood of observing a sample of gene trees given the proposed species network. The sample of gene trees can include branch lengths, in which case the likelihood is derived from the time intervals between successive coalescent events \cite{YuEtAl14}. In the absence of branch lengths, the likelihood is derived from the probability mass of each gene tree topology \cite{YuEtAl12,YuEtAl14}. This probability is marginalized over every coalescent history $h$, which is all the ways for a gene tree to follow the reticulate branching of the network:

\begin{equation}
P(g|\Psi,\Gamma,\theta) = \sum_{h \in H_\Psi (g)}P(h|\Psi,\Gamma,\theta)
\end{equation}

and the ML species network is therefore:

\begin{equation}
\label{eq:like}
\hat{\Psi} = \argmax_\Psi{\prod_{g \in \G}{P(g|\Psi,\Gamma,\theta)}}.
\end{equation}

ML inference of species networks has been implemented as the \texttt{InferNetwork\_ML} command in PhyloNet \cite{YuEtAl14}, which identifies the ML species network up to a maximum number of reticulations.

 Similar to our discussion of the MDC criterion above, absent any explicit stopping criterion or a penalty term in the likelihood function, obtaining an ML estimate according to Eq. \eqref{eq:like} can result in 
 overly complex phylogenetic networks since adding more reticulations often improves the likelihood of the resulting network. Therefore, it is important to parameterize the search by the number of reticulations 
 sought, $m$, and solve 
 
 \begin{equation}
\label{eq:like1}
\hat{\Psi}(m) = \argmax_{\Psi(m)}{\prod_{g \in \G}{P(g|\Psi(m),\Gamma,\theta)}},
\end{equation}
where the value $m$ is experimented with by observing the improvement in the likelihood for varying values of $m$ (for example, maximum likelihood inference of phylogenetic networks 
in PhyloNet implements information criteria, such as AIC and BIC, for this purpose \cite{YuEtAl14}).

The computational cost of the likelihood calculation increases with larger species networks and gene trees. Not only does this increase the number of branches and coalescent times, but as more reticulations are added many more possible coalescent histories exist to be summed over. Even with one reticulation edge attached onto a tree, the difficulty of the problem is exponential to tree cases. The computational complexity of the likelihood calculation is highly related to the size of the set of all coalescent histories of a gene tree conciliated in a network. Zhu {\em et al.} \cite{Zhu289207} proposed an algorithm to compute the number of coalescent histories of a gene tree for a network, and demonstrated that the number can grow exponentially after adding merely one reticulation edge to a species tree.

\begin{figure}[!htp]
\centerline{\includegraphics[width=0.8\textwidth]{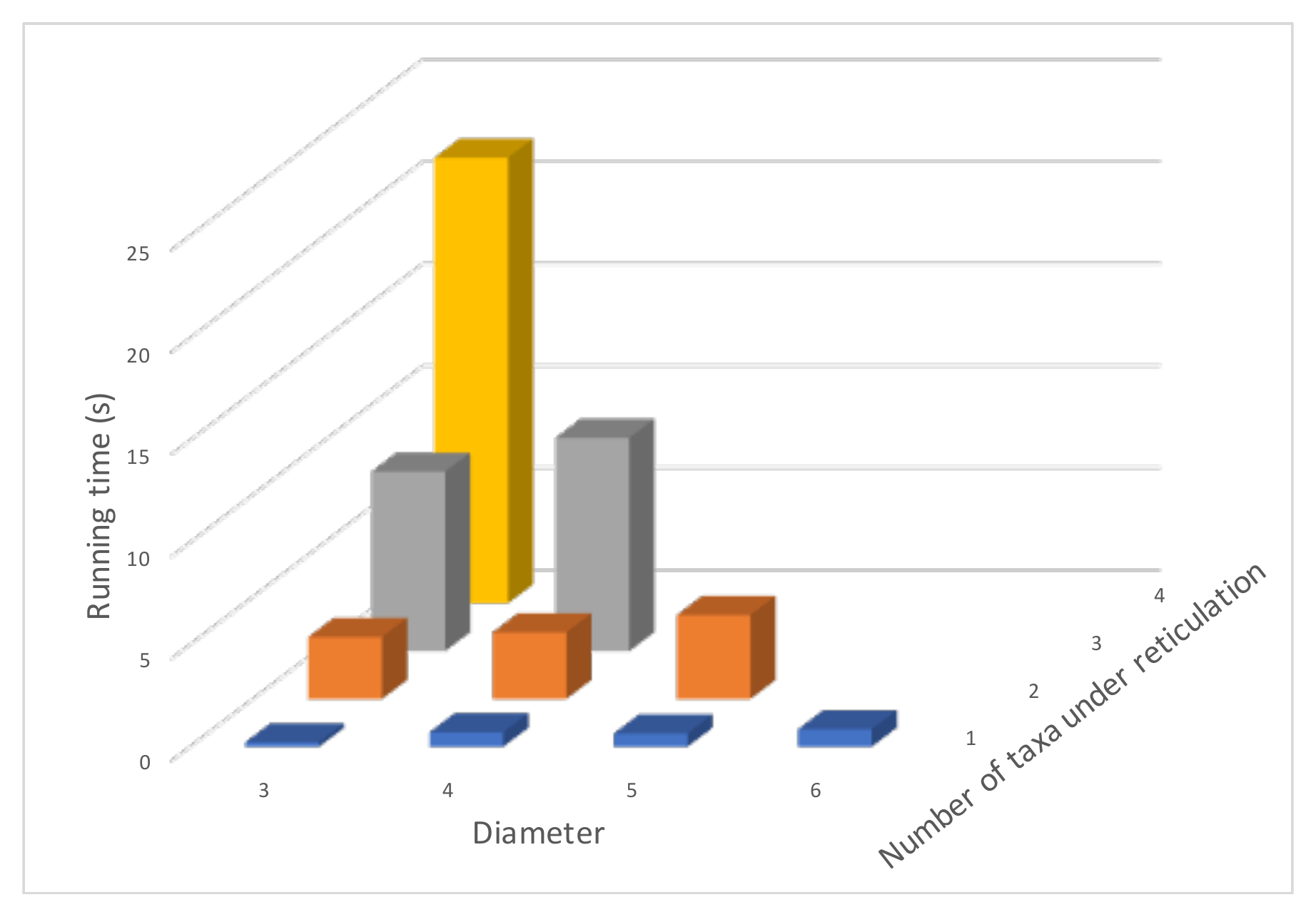}}
\caption{{\bf Running time of computing the likelihood of a phylogenetic network given gene tree topologies.} 150 1-reticulation phylogenetic networks with 5 species and 4 individuals per species was used, and the data consisted of 10,000 bi-allelic markers.  The networks varied in terms of the diameter of a reticulation node (the number of edges on the cycle in the underlying 
undirected graph) and the number of taxa (leaves) under the reticulation nodes. \label{fig:runningTime}}
\end{figure}

To show how running time of likelihood computation varies in a network with a single reticulation, we generated 150 random 1-reticulation networks with 5 taxa, then simulated 10,000  bi-allelic markers with 4 individuals per species. When a reticulation node exists in a phylogenetic network, this will induce a cycle in the unrooted equivalent of the acyclic rooted network. The ``diameter'' is the length of that cycle, and we ran the likelihood computation in \cite{zhu2017bayesian} and summarized the maximum running time according to values of diameter and number of taxa under reticulation. Fig. (\ref{fig:runningTime}) shows that the complexity of the likelihood computation is highly related to the structure of the network. The running time in the worst case is hundreds of times slower than that of the best case.

A faster way to estimate a species network is to calculate a pseudolikelihood instead of a full likelihood. The \texttt{InferNetwork\_MPL} command in PhyloNet implements a maximum pseudo-likelihood (MPL) method for species networks. This method is based on rooted triples, which is akin to the MP-EST method for species tree inference \cite{YuNakhleh15b,Liu2010}.

Unlike phylogenetic trees, a given phylogenetic network is not necessarily uniquely distinguished by its induced set of rooted triples. Therefore this method cannot distinguish the correct network when other networks induce the same sets of rooted triples \cite{YuNakhleh15b}. However, it is much more scalable than ML methods in terms of the number of taxa \cite{Hejase2016}.

\begin{figure}[!htp]
\centerline{\includegraphics[width=0.8\textwidth]{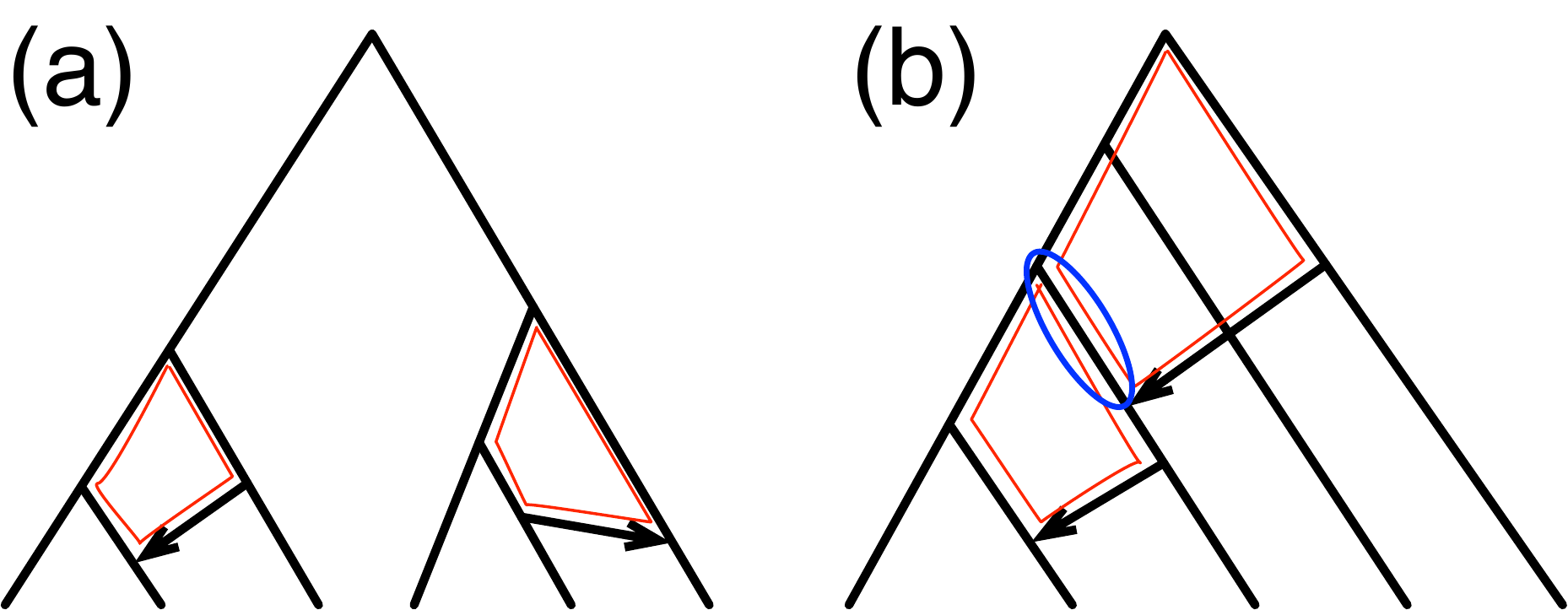}}
\caption{{\bf Level-1 network definition.} Reticulation nodes induce cycles in the (undirected graphs underlying the) phylogenetic networks. The edges of the cycles are highlighted with red lines.
 (a) A level-1 network is one where no edge of the network is shared by two or more cycles. (b) A non-level-1 network is one where at least one edge is shared by at least two cycles (the shared edge in this case is the one inside the blue circle).\label{fig:level1networks}}
\end{figure}

Another MPL method, SNaQ, is available as part of the PhyloNetworks software \cite{solis2016inferring,phylonetworks}. SNaQ is based on unrooted quartets, akin to the ASTRAL method for species tree inference \cite{doi:10.1093/bioinformatics/btu462}. It is even more scalable than \texttt{InferNetwork\_MPL} \cite{Hejase2016}, but can only infer level-1 networks (Fig.~\ref{fig:level1networks}).

\section{Bayesian Inference of Phylogenetic Networks}
Maximum likelihood estimation of phylogenetic networks, as described in the previous section, has three main limitations: 
\begin{itemize}
\item As discussed, without penalizing the complexity of the phylogenetic network, the ML estimate could be an overly complex network 
 with many false reticulations. 
\item The inference results in a single point estimate that does not allow for assessing confidence in the inferred network. 
\item The formulation does not allow for making use of the sequence data directly, but is based on gene tree estimates. 
\end{itemize}
One way of addressing these limitations is to adopt Bayesian inference where an estimate of the posterior distribution on networks is sought directly 
from the sequence data of the individual loci, and where the prior distribution on phylogenetic networks accounts for model complexity in a principled 
manner.  

Before we describe the work on Bayesian inference, it is important to note that while maximum likelihood estimation is not satisfactory, we cannot say that Bayesian estimation is without challenges. Such methods like \cite{wen2017co,zhang2017bayesian,zhu2017bayesian} are based on reversible-jump MCMC \cite{green1995reversible} with varying numbers of parameters. Mixing problems arise when they involve dimension changing moves: adding a reticulation and removing a reticulation. This is because while walking over the space of phylogenetic networks, these methods jump between probability spaces of different models. Therefore moves should be carefully designed to account for mixing issues.

\subsection{Probability Distributions Over Species Networks}
\label{sec:netprior}

It is useful to define probability distributions over species trees or networks without reference to sequence data or gene trees. Among other uses, these probability distributions can be applied as prior distributions in Bayesian inference. The two most common types of prior distributions used for species trees are birth-death tree priors and compound priors. Both types have been extended to create probability distributions over species networks.

As their name implies, birth-death tree priors combine a rate of \textit{birth} with a rate of \textit{death}. These are the rates at which one lineage splits into two, and one lineage ceases to exist respectively \cite{Gernhard:2008uq}. In the context of species trees these rates are more informatively called \textit{speciation} and \textit{extinction}. When the extinction rate is set to zero, this is known as a Yule prior \cite{Yule24}. Birth-death tree priors have been extended to support incomplete sampling in the present, and sampling-through-time \cite{STADLER2010396}. A birth-death prior for species networks has been developed, called the birth-hybridization prior. This prior combines a rate of birth (or speciation) with a rate of hybridization, which is the rate at which two lineages merge into one. This model does not include a rate of extinction \cite{zhang2017bayesian}.

All birth-death tree priors induce a uniform probability distribution over ranked tree topologies, regardless of the rates of speciation and extinction. This means that birth-death priors favor symmetric over asymmetric trees, as symmetric trees have more possible ranked histories. Empirical trees generally have more asymmetric shapes than predicted by birth-death models \cite{doi:10.1093/sysbio/syv001}. For the birth-hybridization prior the probability distribution over network topologies is not invariant to the hybridization rate, which when set to zero reduces to the Yule model, and any topology containing a reticulation will have zero probability.

All birth-death priors are generative, as is the birth-hybridization prior. This means that not only can these distributions be used as priors for Bayesian inference, but they can be used to simulate trees and networks. These simulated distributions can then be used for ABC inference, which is used for models which are difficult to implement using MCMC. They can also be used for posterior predictive checks, which is an absolute measure of model goodness of fit \cite{10.2307/24306036}.

While birth-death priors induce a probability distribution over topologies and branch lengths, compound tree and network priors are constructed from separate distributions on both. Typically compound tree priors combine a uniform distribution over unranked tree topologies, favoring more asymmetric trees. Empirical trees generally have more symmetric shapes than predicted by this distribution \cite{doi:10.1093/sysbio/syv001}. Then a continuous distribution such as gamma can be applied to branch lengths or node heights. Compound priors are used for network inference by adding a third distribution describing the number of reticulations \cite{wen2017co}. A Poisson distribution is a natural fit for this parameter as it describes a probability on non-negative integers. The probability distribution for each network topology can still be uniform for all networks given $k$ reticulations.

Unlike birth-death priors, compound priors are not generative, so it is not straightforward to simulate trees or networks from those distributions. The most obvious way to simulate such trees and networks would be running an existing Markov Chain Monte Carlo sampler without any data, and subsampling states from the chain at a low enough frequency to ensure independence between samples.

\subsection{Sampling the Posterior Distribution}
The ML species network with $k + 1$ reticulations will always have a higher likelihood than the ML network with $k$ reticulations. For this reason, some threshold of significance must be applied to estimate the number of reticulations. This threshold may be arbitrary or it may be theoretically based, for example the Akaike information criterion (AIC) and Bayesian information criterion (BIC) measures of relative fit \cite{10.2307/2533961}.

In contrast, Bayesian methods of species network inference are able to naturally model the probability distribution over species networks including the number of reticulations by using a prior (see Section~\ref{sec:netprior}). In a Bayesian model, the posterior probability of a species network $P(\Psi)$ is proportional to the likelihood of the gene trees $P(G|\Psi,\Gamma,\theta)$, multiplied by the prior on the network and other parameters of the model $P(\Psi,\Gamma,\theta)$, and marginalized over all possible values of $\Gamma$ and $\theta$:

\begin{equation}
\label{eq:BayesGT}
P(\Psi) \propto \iint P(G|\Psi,\theta) \cdot P(\Psi,\Gamma,\theta) \diff \Gamma \diff \theta.
\end{equation}

\begin{figure}[!htp]
\centerline{\includegraphics[width=0.8\textwidth]{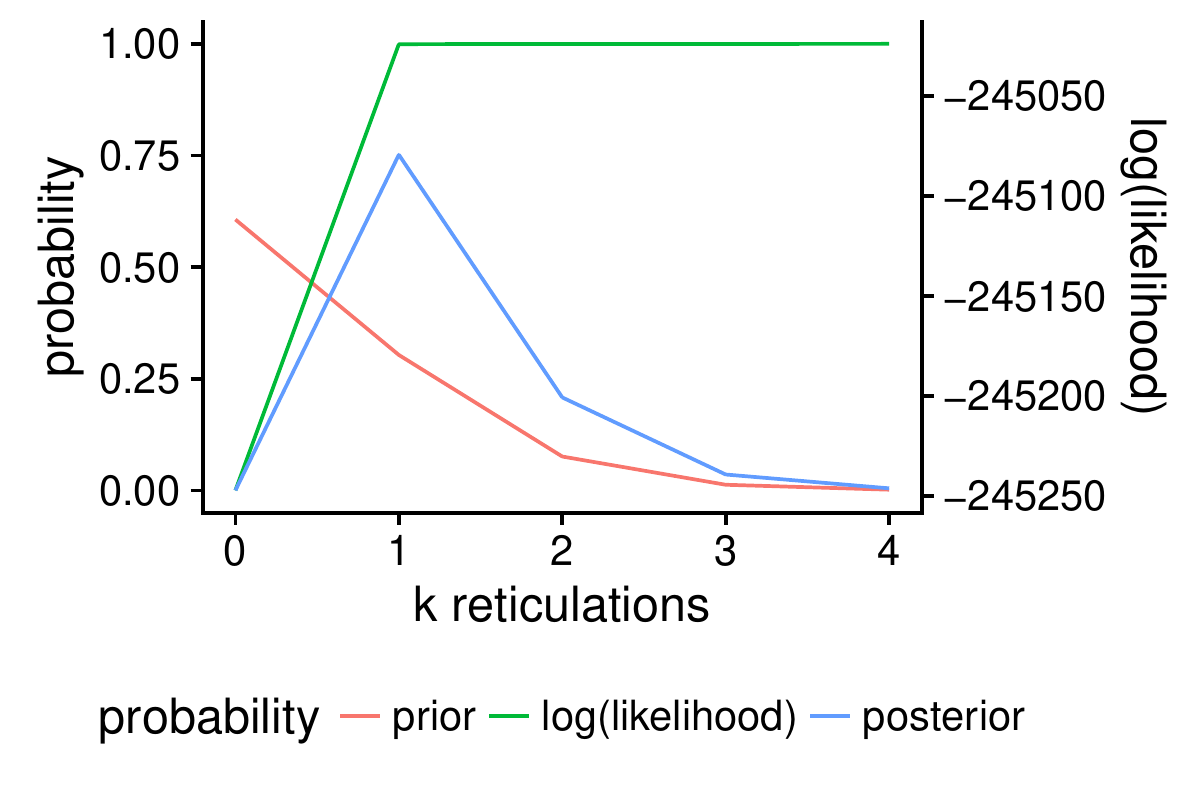}}
\caption{{\bf Bayesian inference of the number of reticulations.} In this example the topology of the network $\Psi$ is fixed, the true number of reticulations is 1, and the likelihood is calculated for a topology with no reticulations, the true reticulation, and additional reticulations, with maximum likelihood branch lengths. The posterior probability was normalized to sum to 1, although as this is not integrated over branch lengths, the typical Bayesian posterior probability might be a bit different. \label{fig:kProbability}}
\end{figure}

When a decaying prior is used on the number of reticulations or on the rate of hybridization, the prior probability of species networks with large numbers of reticulations will be very low, and so will the posterior probability (Fig.~\ref{fig:kProbability}).

Bayesian methods for phylogenetic inference typically use the Metropolis-Hastings Markov Chain Monte Carlo (MCMC) algorithm to estimate the posterior distribution of trees or networks. MCMC is a random walk where each step depends on the previous state, and it is flexible enough to be used for implementing extremely complex models such as species network inference with \textit{relative} ease. Bayesian estimation of species networks from gene trees is implemented in the PhyloNet command \texttt{MCMC\_GT}.

The posterior probability of a species network is equal to the integral in equation~\ref{eq:BayesGT} multiplied by a normalizing constant $Z$ known as the marginal likelihood. In the case of sequential multilocus inference, this constant is equal to $Z = P(G)^{-1}$ The marginal likelihood is usually intractable to calculate, but MCMC sidesteps this calculation by sampling topologies and other parameters with frequencies proportional to their probability mass or density. The posterior probability of a species network $\Psi$ can therefore be approximated as the proportion of steps in the MCMC chain where the network topology $\Psi_i$ at the end of the step $i$ is equal to $\Psi$.

The value of any particular parameter, for example an inheritance probability $\gamma$ for a given reticulation node $v$, can be estimated by averaging its value over the set of $X$ steps where the state includes that parameter. In this case it is averaged over the states where the species network includes that node, i.e. the set $X = \{i: v \in \Psi_i\}$:

\begin{equation}
E(\gamma) = \frac{1}{|X|} \sum^i_{X} \gamma_i
\end{equation}

Bayesian inference has also enabled the inference of species trees and networks directly from multilocus sequence data. Instead of first estimating individual gene trees from multiple sequence alignments, these methods jointly infer the gene trees and species network using an application of Bayes' rule:

\begin{equation}
\label{eq:BayesSeq}
P(G,\Psi) \propto \iint P(D|G) \cdot P(G|\Psi,\Gamma,\theta) \cdot P(\Psi,\Gamma,\theta) \diff \Gamma \diff \theta.
\end{equation}

Here $P(D|G)$ is the likelihood of the data over all gene trees. In practice this is the sum of phylogenetic likelihoods $\sum_i P(d_i|g_i)$ for every sequence alignment $d$ and associated gene tree $g$. As with sequential Bayesian inference of species trees and networks, the use of MCMC avoids the calculation of the marginal likelihood, which for joint inference can be expressed as $Z = P(D)^{-1}$. Joint Bayesian inference was first developed for species trees, and now has several popular implementations including StarBEAST2 \cite{doi:10.1093/molbev/msx126} and BPP \cite{doi:10.1093/sysbio/syw119}.

Joint Bayesian inference of species networks has been implemented independently as the PhyloNet command \texttt{MCMC\_SEQ}, and as the BEAST2 package ``SpeciesNetwork'' \cite{wen2017co,zhang2017bayesian}. These two methods are broadly similar in their model and implementation, with a few notable differences. \texttt{MCMC\_SEQ} uses a compound prior on the species network, whereas SpeciesNetwork has a birth-hybridization prior (see Section~\ref{sec:netprior}). SpeciesNetwork is able to use any of the protein and nucleotide substitution models available in BEAST2. \texttt{MCMC\_SEQ} can be used with any nested GTR model but with fixed rates and base frequencies. So the rates (e.g. the transition/transversion ratio for HKY) must be estimated before running the analysis, or Jukes-Cantor is used where all rates and base frequencies are equal. 


\subsection{Inference Under MSC vs. MSNC When Hybridization Is Present}
\begin{figure}[!htp]
\centerline{\includegraphics[width=1\textwidth]{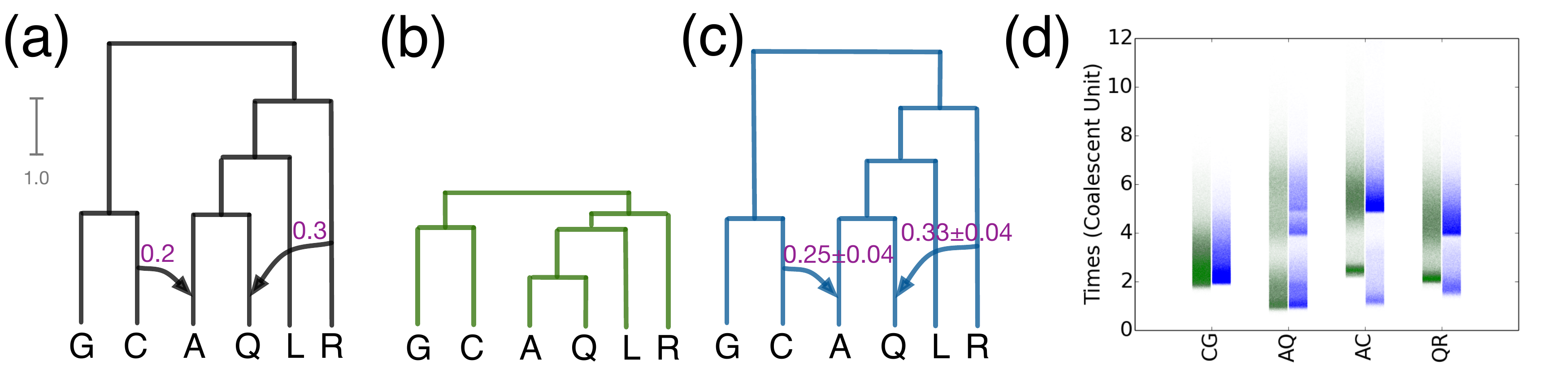}}
\caption{{\bf Inference under the MSC and MSNC when the evolutionary history involves hybridization.} (a) The true phylogenetic network with the shown inheritance probabilities and branch lengths (in coalescent units).  (b) The MPP (maximum a posteriori probability) species tree estimated under the MSC by StarBEAST (frequency of 94\% in the $95\%$ credible set) with the average divergence times. (c) The MPP phylogenetic network along with the inheritance probabilities estimated under the MSNC by \texttt{MCMC\_SEQ} \cite{wen2017co} (the only network topology in the $95\%$ credible set). The scale bar of divergence times represents 1 coalescent unit for (a-c).  (d) The coalescent times of the MRCAs of (C,G), (A,Q), (A,C), (Q,R) from co-estimated gene trees inferred by StarBEAST (green) and \texttt{MCMC\_SEQ} (blue).  \label{fig:simulation}}
\end{figure}

We simulated 128 loci on the phylogenetic network of Fig. \ref{fig:simulation}(a).  The program \texttt{ms} \cite{Hudson02} was used to simulate 128 gene trees on the network, and each gene 
tree was used to simulate a sequence alignment of 500 sites using the program Seq-gen \cite{seq-gen} under the GTR model and $\theta = 0.036$ for the population mutation rate. The exact command used was:
\begin{itemize}
\item[]\texttt{seq-gen -mgtr -s$0.018$ -f$0.2112,0.2888,0.2896,0.2104$ -r$0.2173,\\0.9798,0.2575,0.1038,1,0.2070$ -l$500$}
\end{itemize}
We then ran both StarBEAST and \texttt{MCMC\_SEQ}, as inference methods under the MSC and MSNC models, respectively, for $6\times 10^7$ iterations each. The results are shown in Fig. \ref{fig:simulation}. 

A few observations are in order. First, while StarBEAST is not designed to deal with hybridization, it inferred the tree topology (Fig. \ref{fig:simulation}(b)) that is obtainable by removing the 
  two hybridization events (the two arrows) from the true 
  phylogenetic network (the backbone tree). Second, \texttt{MCMC\_SEQ} identified the true phylogenetic network as the one with the highest posterior (Fig. \ref{fig:simulation}(c)).
  Furthermore, the estimated inheritance
  probabilities are very close to the true ones. Third, and most interestingly, since StarBEAST does not account for hybridization, it accounts for all heterogeneity across loci as being caused by 
  incomplete lineage sorting (ILS) by underestimating all branch lengths (that is, ``squashing" the divergence times so as to explain the heterogeneity by ILS). Indeed, Fig. \ref{fig:simulation}(d) shows that the minimum coalescent times of the co-estimated 
  gene trees by StarBEAST force the divergence times in the inferred species tree to be very low. \texttt{MCMC\_SEQ}, on the other hand, accurately estimates the branch lengths 
  of the inferred phylogenetic network since networks differentiate between divergence and hybridization times. For example, Fig. \ref{fig:simulation}(d) shows that 
  the coalescent times of clade (C,G) across all co-estimated gene trees is a continuum with a minimum value around $2$, which defines the divergence time of these two 
  taxa in the phylogenetic network. \texttt{MCMC\_SEQ} clearly identifies two groups of coalescent times for each of the two clades (A,C) and (Q,R): The lower group of coalescent 
  times correspond to hybridization, while the upper group of coalescent times correspond to the coalescences above the respective MRCAs of the clades. We also 
  note that the minimum value of coalescent times corresponding to (Q,R) is larger than that corresponding to (A,Q), which correctly reflects the fact that hybridization 
  from R to Q happened before hybridization from C to A, as indicated in the true phylogenetic network. Finally, for clade (A,Q), three groups of coalescence times are 
  identified by \texttt{MCMC\_SEQ}, which makes sense since there are three common ancestors of A and Q in the network: at the MRCA of (A,Q) in the case of no hybridization involving 
  either of the two taxa, at the MRCA of (A,Q,L,R) in the case of the hybridization involving Q, and at the root of the network in the case of the hybridization involving A. More thorough 
  analysis and comparison of inferences under the MSC and MSNC can be found in \cite{wen2017co}.
  
  These results illustrate the power of using a phylogenetic network inference method when hybridization is involved. In particular, if hybridization had occurred, and the practitioner did not 
  suspect it and ran StarBEAST instead, they would get wrong inferences. In this case, the errors all have to do with the divergence time estimates. However, the topology of the inferred tree 
  could be wrong as well, depending on the hybridization scenarios. 

\section{Phylogenetic Invariants Methods}
The focus of this chapter up to this point has largely been on the MSC and MSNC models. A parallel effort has been led to detect reticulate evolution by using the notion of 
 {\em phylogenetic invariants} \cite{cavender1987invariants,lake1987rate}. Phylogenetic invariants are polynomial relationships satisfied by frequencies of site 
 patterns at the taxa labeling the leaves of a phylogenetic tree (and given a model of sequence evolution). Invariants that are predictive of particular tree topologies could then
 be used for inferring the tree topology by focusing on the space of site patterns rather than the space of tree topologies \cite{Felsenstein04}. As Felsenstein wrote in his 
 book, ``invariants are worth attention, not for what they do for us now, but what they might lead to in the future." With the availability of whole-genome data and, consequently, 
 the ability to obtain better estimates of site frequencies, the future is here. Indeed, methods like SVDQuartets \cite{SVDquartets} use phylogenetic invariants to estimate 
 species trees under the MSC model. 
 
 A detailed discussion of phylogenetic invariants in general is beyond the scope of this manuscript. 
 Interested readers should consult the excellent exposition on the subject in Felsenstein's seminal book (Chapter 22 in \cite{Felsenstein04}). In this section, we briefly review 
 phylogenetic invariants-based methods for detecting reticulation, starting with the most commonly used one, known as the $D$-statistic or the ``ABBA-BABA" test. 

\begin{figure}[!htp]
\centerline{\includegraphics[width=0.8\textwidth]{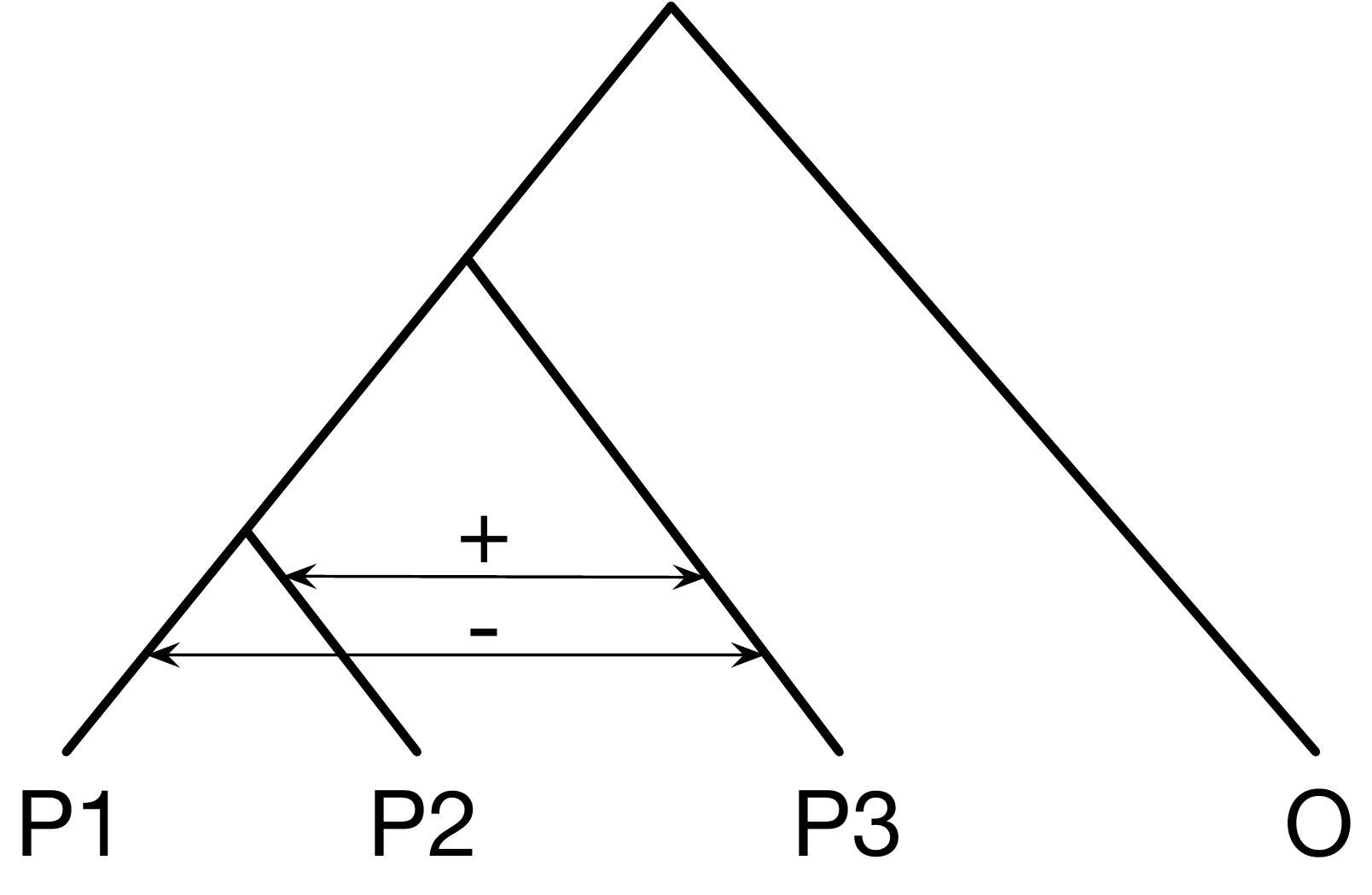}}
\caption{{\bf The four-taxon tree topology used for the $D$-Statistic.} Significant deviations away from a value of 0 of the $D$-statistic (Eq. \eqref{eq:dstat1}) support introgression between P3 and either P1 or P2. As shown, a significant positive value supports introgression between P2 and P3. A significant negative value supports introgression between P1 and P3.\label{fig:dstat1}}
\end{figure}

The $D$-Statistic \cite{Green07052010} is a widely known and frequently applied statistical test for inferring reticulate evolution events. The power of the test to infer reticulate evolution derives from the likelihood calculations of the MSNC. Despite this, the test itself is  simple to calculate and formalize. The $D$-Statistic is given by
\begin{equation}
\label{eq:dstat1}
\frac{N_{ABBA} - N_{BABA}}{N_{ABBA} + N_{BABA}} 
\end{equation}
To calculate these quantities, we are given as input the four taxon tree including outgroup of Fig. \ref{fig:dstat1} and a sequence alignment of the genomes of P1, P2, P3, and O. Given this alignment, $N_{ABBA}$ is calculated as the number of occurrences of single sites in the alignment where P1 and O have the same letter and P2 and P3 have the same letter, but these two letters are not the same i.e. CTTC or GCCG. Similarly, $N_{BABA}$ can be calculated as the number of occurrences in the alignment where the letters of $P1 = P3$ and $P2 = O$ with no other equalities between letters. 

Upon calculating the $D$-Statistic, a significant deviation away from a value of 0 gives evidence for reticulate evolution. As shown in Fig. \ref{fig:dstat1}, a strong positive value implies introgression between P2 and P3 while a strong negative value implies introgression between P1 and P3. No such conclusions can be made from a D value very close to 0. 

The crux of the theory behind the $D$-Statistic lies in the expectation of the probabilities of discordant gene trees given the overall phylogeny of Fig. \ref{fig:dstat1}. If we remove the two reticulation events in Fig. \ref{fig:dstat1} we end up with a species tree $\Psi$. 
Given $\Psi$, the two gene trees whose topologies disagree with that of the species tree are equally probable 
 under the MSC.  
\begin{figure}[!htp]
\centerline{\includegraphics[width=1.0\textwidth]{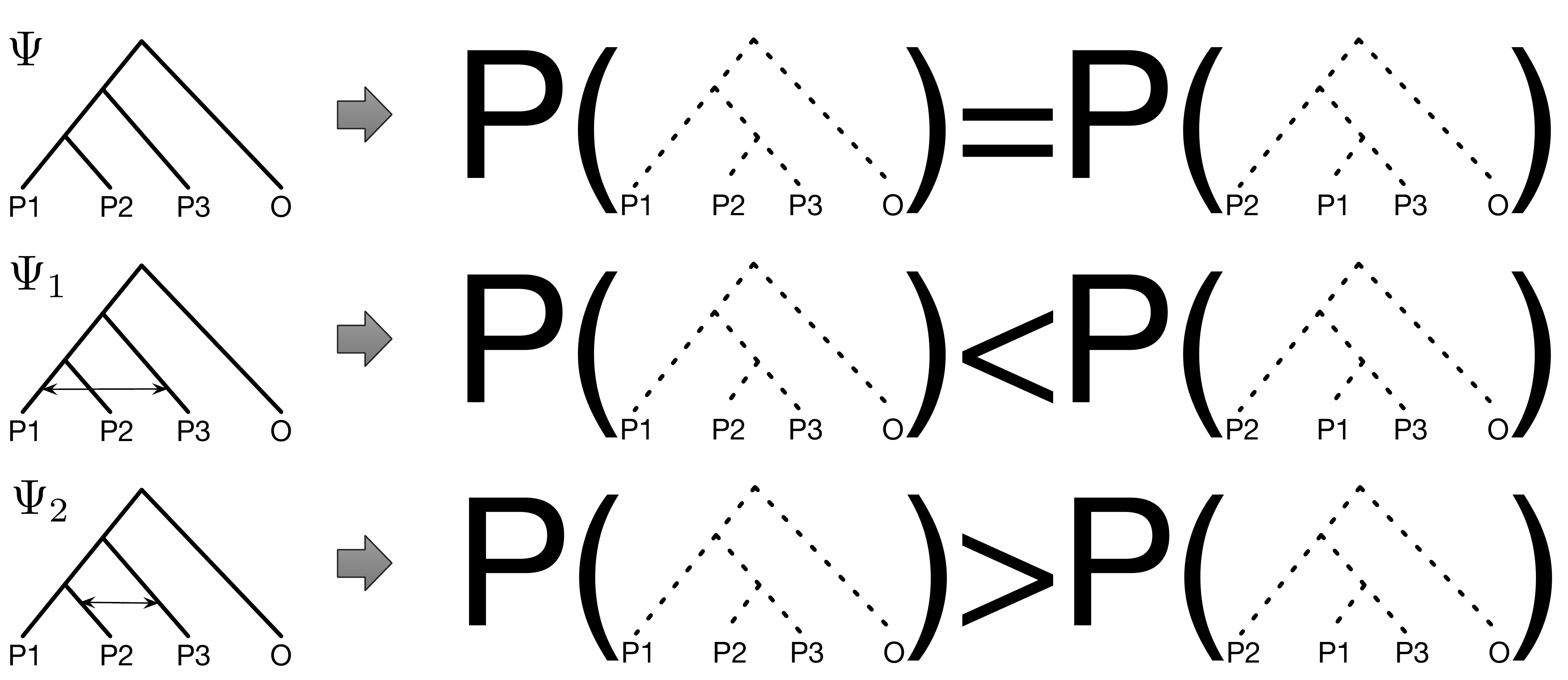}}
\caption{{\bf The three scenarios of probabilities of the two gene trees that are discordant with the species tree in the case of a single reticulation event between P3 and one of the other 
two (in-group) species.} If the evolutionary history of the species is a tree $(\Psi)$, the two discordant gene trees are equally probable. However, if the evolutionary history of the species 
is non-treelike, as given by phylogenetic networks $\Psi_1$ and $\Psi_2$, then the probabilities of the two discordant gene trees are unequal in different ways.\label{fig:dstatgts}}
\end{figure}
 On the other hand, when a reticulation between P1 and P3 occurs, this results in an increase in the probability of the discordant gene tree that groups P1 and P3 as sister taxa, as compared 
 to the other discordant gene tree. Similarly, when a reticulation between P2 and P3 occurs, this results in an increase in the probability of the discordant gene tree that groups P2 and P3 
 as sister taxa. These three scenarios are illustrated in Fig. \ref{fig:dstatgts}. 
 
Assuming an infinite sites model of sequence evolution, the frequencies of gene trees ((P1,(P2,P3)),O) and (((P1,P3),P2),O) directly correlate with the values  $N_{ABBA}$ and $N_{BABA}$,
 respectively, explaining the rationale behind Eq. \eqref{eq:dstat1}. To apply the $D$-statistic, frequencies of the $ABBA$ and $BABA$ site patterns are counted across an alignment of four 
 genomes, the value of Eq. \eqref{eq:dstat1} is calculated, and deviation from $0$ is assessed for statistical significance. A significant deviation is taken as evidence of introgression.  

Since the introduction of the $D$-Statistic, work has been done to extend this framework. Recently, the software package HyDe \cite{blischak2018hyde} was introduced with several extensions including handling multiple individuals from four populations as well as identifying individual hybrids in a population based on the method of \cite{kubatko2015invariants}. In HyDe, higher numbers of individuals are handled through calculating statistics on all permutations of quartets of the individuals. Another recent extension to move the $D$-Statistic beyond four taxa is the D\textsubscript{FOIL} framework introduced by \cite{pease2015detection}. In it, we see the same derivation used in the $D$-Statistic on a particular five-taxon tree. This derivation includes isolating gene trees whose probabilities go from equal to unequal when going from the tree case to the network case as well as converting these gene trees to corresponding site patterns to count in an alignment. Finally, Elworth {\em et al.} recently devised a heuristic, $D_{GEN}$, for automatically deriving phylogenetic invariants for detecting 
hybridization in more general cases than the $D$-Statistic and D\textsubscript{FOIL} can handle \cite{dgen}. The rationale behind the approach of Elworth {\em et al.} is that invariants 
could be derived by computing the probabilities of gene trees under a given species tree (e.g., using the method of \cite{DegnanSalter05}) and then computing the probabilities of the same 
trees under the same species tree with any reticulation scenarios added to it (using the method of \cite{YuEtAl12}), and contrasting the two to identify sets of gene trees whose equal 
probabilities under the tree model get violated under the network model.

The $D$-Statistic is very simple to implement and understand, and it can be calculated 
 on 4-genome alignments very efficiently, making it an appealing choice of a test for detecting introgression. Indeed, applications of the $D$-Statistic are widespread in the literature, reporting on introgression in ancient hominids \cite{Green07052010,Durand01082011}, butterflies \cite{Heliconius2012}, 
 bears   \cite{kumar2017evolutionary}, and  sparrows \cite{elgvin2017genomic}, just to name a few. However, it is important to note here that the derivation of the $D$-Statistic (and its extensions) relies on many 
 assumptions that can easily be violated in practice. One major cause of such a violation is that the mathematics behind the $D$-Statistic relies on the coalescent which makes many simplifying 
 assumptions about the evolutionary model and processes taking place. Of course, this shortcoming applies as well to all phylogenetic inference methods that employ the MSC or MSNC models.
  A second cause of such a violation is that in practice, more than a single reticulation could have taken place and 
 ignoring those could result in erroneous inferences \cite{dgen}. A third violation stems from the way the $D$-Statistic is applied.\footnote{We especially thank David Morrison for requesting that 
 we highlight this issue.} In propositional logic, the statement ``If $p$, then $q$" and 
 its converse ``If $q$, then $p$" are logically not equivalent. That is, if one is true, it is not necessarily the case the other is. Looking back at Fig. \ref{fig:dstatgts},  the 
 statement illustrated by the figure is: If there is no reticulation (i.e., the species phylogeny is a tree), then the probabilities of the two discordant trees are equal. The converse (if the probabilities 
 of the two discordant trees are equal, then the species phylogeny is a tree) does not follow logically. However, this is how the test is used in practice. In all fairness, though, this logical 
 fallacy is commonplace in inferences in biology, including when inferring species trees and networks under the MSC and MSNC, respectively. The fallacy is always dealt with by resorting 
 to the ``simplest possible explanation" argument. For example, why various scenarios could have given rise to equal frequencies of the frequencies of the $ABBA$ and $BABA$ site patterns, 
 the species tree scenario is considered the simplest such possible explanation and is invoked as such. 
 
 Last but not least, Peter \cite{peter2016admixture} recently provided a review and elegant connections between the $D$-Statistic and a family of statistics known as the $F$-statistics.



%

\section{Phylogenetic Networks in the Population Genetics Community}
 The population genetics community has long adopted rooted, directed acyclic graphs as a model of evolutionary histories,
 typically of individuals within a single species. Ancestral recombination graphs, or ARGs, were introduced  \cite{hudson1990gene,GriffithsMarjoram96} 
 to model the evolutionary history of a set of genomic sequences in terms of the coalescence and recombination events that occurred since their 
 most recent common ancestor. Statistical methods for inference of ARGs from genome-wide data have also been developed \cite{rasmussen2014genome}. 
 Gusfield's recent book \cite{gusfield2014recombinatorics} discusses algorithmic and combinatorial aspects of ARGs. However, while ARGs take the 
 shape of a phylogenetic network as defined above, they are aimed at modeling recombination and methods for their inference are generally not applicable 
 to hybridization detection. 
 
 Efforts in the population genetics community that are aimed at modeling admixture and gene flow are more relevant to hybridization detection. Here 
 we discuss one of the most popular methods in this domain, namely TreeMix \cite{pickrell2012inference}. In population genetics, the counterparts to species trees and phylogenetic 
 networks are population trees and admixture graphs, respectively. The difference between these models boils down to what labels their leaves: If the leaves 
 are labeled by different species, then the models are called species trees/networks, and if the leaves are labeled by different sub-populations of the same 
 species, then the models are called population trees and admixture graphs. Of course it is not always easy to identify whether a species or sub-populations have been delimited, and hence what particular tree/network should be called in that case, as species and populations may exist on a continuum \cite{deQueiroz2007sysbio}.

\subsection{TreeMix}
 TreeMix \cite{pickrell2012inference} models the evolution of a set of SNPs where the input data consists of allele frequencies of these SNPs in a set of 
 populations whose evolutionary history is given by a population tree (in the case of no migration) or an admixture graph (in the case where migration is included).

The basis of the model used in TreeMix is in the notion of modeling drift over time as a diffusion process, where an original allele frequency of $x_1$ of a given SNP undergoes drift by an amount $c$ to give rise to a new allele frequency $x_2$ \cite{cavalli1967phylogenetic,nicholson2002assessing,coop2010using}, as given by  
\begin{equation}
\label{eq:treemix1}
x_2 = x_1 + N(0,c \cdot x_1[1-x_1]).
\end{equation}
 It is worth noting here that, as pointed out in \cite{pickrell2012inference}, $c ~= t/2N_e$ for drift over small time scales where the time scale is on the same order of the effective population size \cite{nicholson2002assessing}.

When there are multiple populations under the effects of drift that evolved down a tree, the drift processes become linked and can no longer be described with independent Gaussian additions. This process is modeled with a covariance matrix derived from the amounts of drift occurring along the branches of the evolutionary tree. Finally, to incorporate reticulate evolution into the model one only needs to alter this covariance matrix based on the rate of gene flow along reticulate edges in the admixture graph. 

In its current implementation, the authors of TreeMix assume the evolutionary history of the sampled, extant populations is very close to a tree-like structure. Based on this assumption, the 
search for a maximum likelihood admixture graph proceeds by first estimating a rooted tree, and then adding migration events one at a time until they are no longer statistically significant (however, 
as the authors point out, they ``prefer to stop adding migration events well before this point so that the result graph remains interpretable."). Clearly, adding additional edges connecting the edges of a tree in this way will infer a tree-based network, which is a more limited class of networks compared with phylogenetic networks \cite{francis2015phylogenetic}.

%



\section{Data, Methods, and Software}

Given the interest in reticulate evolution from both theoreticians and empirical researchers, it is perhaps unsurprising that software to infer hybridization has proliferated in recent years. Such methods have been developed for a variety of data types, including multilocus data, SNP matrices, and whole genomes (Table~\ref{tab:inferMethods}). Some of these methods are able to infer a phylogenetic network, whereas others infer introgression between species tree lineages. With the exception of TreeMix and MixMapper, methods which infer networks are not constructed around a backbone tree, and so do not assume that tree-like evolution is the dominant process.

Regardless of the input and output, most of these methods allow for ILS in addition to hybridization (Table~\ref{tab:inferMethods}), which is necessary to infer phylogenetic networks representing reticulate evolution in biological systems where ILS is a possibility. Likelihood (including Bayesian) methods incorporate the possibility or effect of ILS into the likelihood function. Maximum parsimony methods that minimize deep coalescences, for example \texttt{InferNetwork\_MP}, essentially attempt to infer the tree or network that minimizes the quantity of ILS, but do not necessarily eliminate all genetic discordance.

Methods which do not allow for ILS will instead infer phylogenetic networks representing conflicting signals \cite{doi:10.1093/molbev/msh018}. Reticulation is one such conflicting signal, but so is ILS, so reticulate branches in these networks should not be blindly interpreted as necessarily representing introgression or hybridization.

\begin{sidewaystable}
\vspace{4in}
\hspace*{-3cm}\caption{Common methods to infer hybridization}
\label{tab:inferMethods}
\hspace*{-1.8cm}\begin{tabular}{l|lllll}
\hline
 Method                                        & Platform                           & Inference               & ILS       & Input                                  & Output                                                     \\
\hline
 5-taxon ABBA-BABA \cite{pease2015detection}   & $D_{FOIL}$                         & Phylogenetic invariants & Yes       & Genomic data                           & Presence/absence of introgression\footnote{along with statistical significance}     \\
 ABBA-BABA \cite{Neandertal10}                 & $D_{FOIL}$, HyDe, etc.             & Phylogenetic invariants & Yes       & Genomic data                           & Presence/absence of introgression$^a$               \\
  $D_{GEN}$ \cite{dgen}                         & ALPHA \cite{elworth2018alpha}      & Phylogenetic invariants & Yes       & Genomic data                           & Presence/absence of introgression$^a$     \\
 Blischak et al. \cite{blischak2018hyde}       & HyDe                               & Phylogenetic invariants & Yes       & Genomic data                           & Hybrid species   \\
 AIM \cite{Mueller348391}                      & BEAST2                             & Bayesian                & Yes       & Multilocus sequences                   & Rooted tree w/ gene flow\footnote{between contemporaneous lineages}\\
 IMa2 \cite{doi:10.1093/molbev/msp296}         & Standalone                         & Bayesian                & Yes       & Multilocus sequences and backbone tree & Evolutionary parameters\footnote{effective population sizes, migration rates, divergence times}            \\
 PIRN \cite{wu2013algorithm}         & Standalone                         & Maximum parsimony       & No        & Rooted gene trees                      & Rooted network                                             \\
 CASS \cite{van2010phylogenetic}                              & Dendroscope \cite{dendroscope}     & Maximum parsimony       & No        & Rooted gene trees                      & Rooted network                                             \\
 InferNetwork\_ML \cite{YuEtAl14}              & PhyloNet \cite{phylonet,wen2018inferring}           & Maximum likelihood      & Yes       & Rooted gene trees                      & Rooted network                                             \\
 InferNetwork\_MP \cite{YuEtAl13}              & PhyloNet \cite{phylonet,wen2018inferring}           & Maximum parsimony       & Minimized & Rooted gene trees                      & Rooted network                                             \\
 InferNetwork\_MPL \cite{YuNakhleh15b}         & PhyloNet \cite{phylonet,wen2018inferring}           & Pseudolikelihood        & Yes       & Rooted gene trees                      & Rooted network                                             \\
 MCMC\_BiMarkers \cite{zhu2017bayesian}        & PhyloNet \cite{phylonet,wen2018inferring}           & Bayesian                & Yes       & Biallelic sites                        & Rooted network                                             \\
 MCMC\_GT \cite{WenEtAl16a}                    & PhyloNet \cite{phylonet,wen2018inferring}           & Bayesian                & Yes       & Rooted gene trees                      & Rooted network                                             \\
 MCMC\_SEQ \cite{wen2017co}                    & PhyloNet \cite{phylonet,wen2018inferring}           & Bayesian                & Yes       & Multilocus sequences                   & Rooted network                                             \\
 MLE\_BiMarkers \cite{Zhu289207}               & PhyloNet \cite{phylonet,wen2018inferring}           & Maximum (pseudo)likelihood & Yes    & Biallelic sites                        & Rooted network                                             \\
 MixMapper \cite{mixmapper}                    & Standalone                         & Pseudolikelihood        & Yes       & Allele frequencies                     & Rooted network                                             \\
 Neighbor-net \cite{doi:10.1093/molbev/msh018} & Splitstree \cite{splitstree}       & Agglomeration           & No        & Genomic data                           & Splits graph (unrooted network)                            \\
 SNAQ \cite{solis2016inferring}                & PhyloNetworks \cite{phylonetworks} & Pseudolikelihood        & Yes       & Unrooted gene trees                    & Unrooted, level-1 network                                  \\
 SpeciesNetwork \cite{zhang2017bayesian}       & BEAST2 \cite{beast2}               & Bayesian                & Yes       & Multilocus sequences                   & Rooted network                                             \\
 STEM-hy \cite{Kubatko09}                      & Standalone                         & Maximum likelihood      & Yes       & Molecular clock gene trees             & Rooted tree w/ hybridizations\footnote{hybridizations limited to sister lineages} \\
 TreeMix \cite{pickrell2012inference}          & Standalone                         & Pseudolikelihood        & Yes       & Allele frequencies                     & User-rooted, tree-based network                            \\
\hline
\end{tabular} \\
\end{sidewaystable}

\subsection{Limitations}

The biggest limitation of methods to infer introgression and hybridization, including species network methods, is scalability.

Methods which infer a species network directly from multilocus sequences have only been used with a handful of taxa, and less than 200 loci. A systematic study of the species tree method StarBEAST found that the number of loci used has a power law relationship with a large exponent with the required computational time, making inference using thousands of loci intractable \cite{Ogilvie01052016}. Although no systematic study of computational performance has been conducted for equivalent species network methods such as \texttt{MCMC\_SEQ}, anecdotally they suffer from similar scaling issues.

Methods which scale better than direct multilocus inference have been developed, but they are no silver bullet. Species networks can be estimated directly from unlinked biallelic markers by integrating over all possible gene trees for each marker, which avoids having to sequentially or jointly estimate gene trees. Biallelic methods make the inference of species trees and networks from thousands of markers possible, at the cost of using less informative markers.

Pseudolikelihood inference has been developed for both biallelic and multilocus methods \cite{Zhu289207,YuNakhleh15b}. This reduces the computational cost of computing the likelihood of a species network as the number of taxa increases, and enabled the reanalysis of an empirical data set with 1070 genes from 23 species \cite{YuNakhleh15b}.

The ABBA-BABA test and similar phylogenetic invariant methods are capable of analyzing an enormous depth of data (whole genomes), but can be limited in taxonomic breadth based on hard limits of four or five taxa for the D-Statistic and $D_{FOIL}$, respectively, or by computational requirements for the case of $D_{GEN}$ (Table \ref{tab:inferMethods}). In addition, the D-Statistic and $D_{FOIL}$ are limited to testing a specific hypothesis for introgression given a fixed species tree topology of a specific shape. This can be understood as a trade off, where the flexibility of species network methods is sacrificed for the ability to use more data.

Beyond scalability, another present limitation is visualizing or summarizing posterior or bootstrap distributions of networks. Methods have been developed to visualize whole distributions of trees, or summarize a distribution as a single tree. Equivalent tools for networks are underdeveloped, leaving researchers to report the topology or set of topologies with the highest posterior or bootstrap support.

\section{Conclusions and Future Directions}

Great strides have been made over the past decade in the inference of evolutionary histories in the presence of hybridization and other processes, most notably incomplete lineage sorting. Species networks can now be inferred directly from species-level data which do not assume any kind of backbone tree, and instead put reticulate evolution on an equal basis with speciation. 

To some extent the development of species network methods have recapitulated the development of species tree methods, starting with maximum parsimony and transitioning to likelihood methods, both maximum likelihood and Bayesian. To improve computational performance and enable the analysis of large data sets, pseudo-likelihood species network methods have been developed, inspired by similar species tree methods. 

Phylogenetic invariant methods such as the ABBA-BABA test are able to test for reticulate evolution across whole genomes, uncovering chromosomal inversions and other features associated with hybridization and introgression. Last but not least, the population genetics community has long been interested in and developing methods for phylogenetic networks mainly to model the evolution of sub-populations in the presence of admixture and gene flow. In this chapter, we surveyed the recent computational developments in the field and listed computer software programs that enable reticulate evolutionary analyses for the study of hybridization and introgression, and generally to infer more accurate evolutionary histories of genes and species.

Empirical biologists feel constrained by the computational performance of existing species network methods. For species trees, phylogenetic invariant methods can be combined with quartet reconciliation to infer large species trees from genomic data, as in SVDquartets \cite{SVDquartets}. For networks, phylogenetic invariant methods to identify the true network with a limited number of edges need to be developed, as do methods to reconcile the resulting subnets.

Even for species trees, Bayesian methods have practical limitations in terms of the amount of data they can be used with. Bayesian methods for trees and networks, with few exceptions, have been built on Markov chain Monte Carlo (MCMC). This technique is inherently serial and hence unsuited to modern workstations, which contain many CPU and GPU cores working in parallel. It is important to continue to explore other Bayesian algorithms which work in parallel such as sequential Monte Carlo \cite{doi:10.1093/sysbio/syr131}, or algorithms which are orders of magnitude faster than MCMC such as variational Bayes \cite{waterhouse1996bayesian}.

Phylogenetic methods for species tree inference have a huge head start on methods for species network inference. Not only is the problem of species network inference much more complicated, but species tree methods have been in development for much longer. For example, MDC for species trees was first described in 1997, and extended to phylogenetic networks 14 years later \cite{maddison,YuThan11}. In this light the progress made is remarkable. However, as evolutionary biology is moving towards data sets containing whole genomes for hundreds or even thousands of taxa, methods developers must focus on improving the scalability of their methods without sacrificing accuracy so that the full potential of this data may be realized.

While preliminary studies exist of the performance of the different methods for phylogenetic network inference \cite{kamneva2017simulation}, more thorough studies are needed to assess 
the accuracy as well as computational requirements of the different methods. 

Last but not least, it is important to highlight that all the development described above excludes processes such as gene duplication and loss, and so may be susceptible to errors and artifacts which can be present in data such as hidden paralogy. Furthermore, the multispecies network coalescent already has its own population-genetic assumptions, almost all of which are 
not necessarily realistic for analyses in practice. Accounting for these 
is a major next step (though it is important to point out that these have not been fully explored in the context of species tree inference either), but the mathematical complexity will most likely add, extensively, to the 
computational complexity of the inference step. 


\begin{acknowledgement}
Luay Nakhleh started working on phylogenetic networks in 2002 in a close collaboration with Tandy Warnow, Bernard M.E. Moret, and C. Randal Linder. At the time, their focus was 
on phylogenetic networks in terms of displaying trees. This focus led to work on inference of smallest phylogenetic networks that display a given set of trees, as well as on comparing 
networks in terms of their displayed trees. While the approaches pursued at the time were basic, they were foundational in terms of pursuing more sophisticated models and approaches 
by Nakhleh and his group. Therefore, we would like to acknowledge the role that Bernard played in the early days of (explicit) phylogenetic networks. 

The authors would also like to acknowledge James Mallet, Craig Moritz, David Morrison, and Mike Steel for extensive discussions and detailed comments that helped us significantly improve this 
chapter. The authors thank Matthew Hahn and Kelley Harris for their discussion of the definition of phylogenetic invariant methods. 

This work was partially supported by NSF grants DBI-1355998, CCF-1302179, CCF-1514177, CCF-1800723, and DMS-1547433.
\end{acknowledgement}


\printindex
\end{document}